\documentclass[superscriptaddress, aps, showpacs, amsmath, amssymb, nofootinbib,notitlepage]{revtex4-1}
\usepackage[T1]{fontenc}
\usepackage[latin9]{inputenc}
\usepackage[a4paper]{geometry}
\usepackage{amstext}
\usepackage{esint}

\usepackage{graphicx}

\usepackage{bm}
\usepackage{braket}

\newcommand{\average}[1]{\left\langle#1\right\rangle}

\begin{document}

\title{Irreversible Work versus Fidelity Susceptibility for infinitesimal quenches}
\author{Simone Paganelli}
\affiliation{Dipartimento di Scienze Fisiche e Chimiche, Universit\`{a} dell'Aquila, via Vetoio, I-67010 Coppito-L'Aquila, Italy}
\author{Tony J. G.  Apollaro}
\affiliation{NEST, Istituto Nanoscienze-CNR and Dipartimento di Fisica e Chimica, Universit$\grave{a}$  degli Studi di Palermo, via Archirafi 36, I-90123 Palermo, Italy}
\affiliation{Dipartimento di Fisica, Università degli Studi di Milano, 20133 Milano, Italy}
\begin{abstract}
We compare  the irreversible work produced in an infinitesimal sudden quench of a quantum system at zero temperature with its ground state fidelity susceptibility, giving an explicit relation between the two quantities. We find that the former  is  proportional to the latter but for an extra term appearing in  the irreversible work which includes also contributions from the excited states.
We calculate explicitly the two quantities in the case of the quantum Ising chain,  showing that at criticality they exhibit different scaling behavior. The irreversible work, rescaled by square of the quench's amplitude, exhibits a divergence slower than the fidelity susceptibility one. As a consequence, the two quantities obey also different finite-size scaling relations.
\end{abstract}
\maketitle

\section{Introduction}

One of the main goals in the field of quantum thermodynamics is the analysis of the statistics of work and heat produced by a quantum system brought out of equilibrium by a change of its dynamical parameters. This type of problems are currently under active investigation as they both address fundamental issues and may pave the way to novel applications of quantum devices. The heat-to-work transformation in a quantum system is crucial for quantum engines implementations \cite{Goold2016, Kosloff2016} and the increasing control achieved in the laboratories over the parameters' tuning of mesoscopic systems \cite{BlochNat12} provides further motivations.
The dynamics of out-of-equilibrium closed systems \cite{Polkovnikov2011} gives the possibility to investigate how fundamental processes occurring in nature, e.g.,  thermalisation, do translate in the realm of closed quantum systems. Moreover,
by means of the fluctuation theorems~\cite{Jarzynski1997,Jarzynski2011,Campisi2011}, which have been tested also experimentally \cite{Toyabe2010, An2014, 2016arXiv160206958B,PhysRevLett.113.140601, PhysRevLett.110.230601, PhysRevLett.115.190601}, it is possible to relate  quantities in an out-of equilibrium protocol - such as work production - to equilibrium quantities, namely free-energy difference.

The typical scenario consists in a quantum system described by an Hamiltonian having a parameter that can be fine-tuned and time-controlled. Quenching such a parameter drives the  system out of the initial equilibrium state, resulting in an irreversible entropy and work production \cite{Gambassi2011,Plastina2014,Fusco2014}. These provide a measure of the irreversibility of the transformation
and have been at the center of many investigations \cite{Talkner2007a,Silva2008,Dorner2012,Gambassi2012,Apollaro2015,Sindona2014,Apollaro2016}. In particular, recent studies have focused on the irreversible work and entropy production of quenches on many-body systems undergoing a quantum phase transition (QPT)  \cite{Dorner2012, Mascarenhas2014,Silva2008,Bayat2016} showing that its critical point (CP)
can be signalled by these out-of-equilibrium quantities. In particular, in a first-order QPT the derivative of the average work is discontinuous at the CP, while in a second-order QPT the derivative of the irreversible entropy production and the variance of the work are discontinuous at its CP.
Furthermore, given an Hamiltonian's parameter $\lambda$, for an infinitesimal sudden quench  $\lambda \rightarrow \lambda+\delta$ crossing the CP, it has been shown in Ref.~\cite{Bayat2016} that the rescaled irreversible work (RIW) $\left\langle \tilde{W}_{irr} \right\rangle=\left\langle W_{irr} \right\rangle_{irr}/\delta^2$, exhibits finite-size scaling at criticality in a two-impurity Kondo model.

At zero temperature, a quench through a CP induces a dramatic change in the ground state of the system. This consideration inspired several works in which QPTs have been studied by the non analyticity of the fidelity between the ground states before and after the quench  \cite{Quan2006,Zanardi2006,You2007,Campos2007}. 
In particular, the scaling behavior of the fidelity susceptibility (FS) $\chi_F$ after an infinitesimal quench, has been shown to be related to the universality classes of the QPT \cite{Gu2010,Gu2008a,Gu2011}.

The properties of  the RIW for infinitesimal sudden quenches resemble the one of a susceptibility, as suggested in \cite{ Mascarenhas2014}. So, at a first sight,  one could be temped to identify RIW and FS as equivalent quantities. 
The scope of this paper is to find a functional relation between $\tilde{W}_{irr}$ and $\chi_F$. We show that the two quantities are not equivalent having the  $\tilde{W}_{irr}$  an extra term taking into account all the transitions to the excited states. Then we analyze the different behavior of RIW and the FS  in the thermodynamical limit for the specific case of a quantum Ising model.
We show that, while for the FS it is possible to define a critical adiabatic dimension and a critical exponent \cite{ Gu2010,Gu2008a,Gu2011}, the RIW which diverges only logarithmically at the critical point.

The paper is organised as follows: in Sec.~\ref{S.Fidelity} and \ref{S.Rel} we report the perturbative expressions for the fidelity susceptibility and the irreversible work, respectively, deriving a functional relation between them. In Sec.~\ref{S.Irr} we compare these two quantities both in the thermodynamical limit and for finite size systems in the quantum Ising model and, finally, in Sec.~\ref{S.Concl} we draw our conclusions.

\section{Fidelity}\label{S.Fidelity}
The fidelity between two pure state is defined by the modulus of their
overlap~\cite{Uhlmann2011}
\begin{equation}
F=\left|\left\langle \psi|\psi'\right\rangle \right|,
\end{equation}
providing a measure of their closeness in the
Hilbert space. The fidelity ranges from zero, for orthogonal states, to one, for identical states. Because at the CP the ground state of a many body system is excepted to undergo a qualitative change, it has been argued that a  QPT can be investigated by the fidelity~\cite{ZanardiPRE06,Gu2010}. 

Given an Hamiltonian dependent on a paramenter $\lambda$
\begin{equation}
H(\lambda)=H_{0}+\lambda H_{1},
\end{equation}
we denote with $E_{n}(\lambda)$ and $\ket{\psi_{n}(\lambda)}$ the
energies and the corresponding eigenstates of $H$ for a given $\lambda$.
The change of the parameter from $\lambda$ to $\lambda'$ modifies
the ground state of the system by an amount that can be quantified
by the fidelity between the two corresponding ground states 
\begin{equation}
F(\lambda,\lambda')=\left|\braket{\psi_{0}(\lambda)|\psi_{0}(\lambda')}\right|.
\end{equation}

For an infinitesimal change in $\lambda$, supposing that the ground-state
wavefunction is differentiable in the parameter space, one has \cite{Gu2010}
\begin{equation}
F(\lambda,\lambda+\delta)=1-\frac{\delta{}^{2}}{2}\chi_{F}(\lambda),
\end{equation}
where $\chi_{F}(\lambda)$ is the fidelity susceptibility which can be written in the differential form as
\begin{equation}
\chi_{F}=\braket{\frac{\partial}{\partial\lambda}\psi_{0}(\lambda)|\frac{\partial}{\partial\lambda}\psi_{0}(\lambda)}-\left|\braket{\frac{\partial}{\partial\lambda}\psi_{0}(\lambda)|\psi_{0}(\lambda)}\right|^{2}.\label{eq:chiF}
\end{equation}
Finally, exploiting perturbation theory, Eq.~\ref{eq:chiF} can be expressed as~\cite{PhysRevLett.99.100603}
\begin{equation}\label{eqn:pertchi}
\chi_{F}(\lambda)=\sum_{n\neq0}\frac{\left|\bra{\psi_{n}(\lambda)}H_{1}\ket{\psi_{0}(\lambda)}\right|^{2}}{\left[E_{n}(\lambda)-E_{0}(\lambda)\right]^{2}}~,
\end{equation}
relating the structural difference of the two ground states to the low-lying energy spectrum.

\section{Irreversible work at $T=0$ and relation to the fidelity susceptibility}\label{S.Rel}

We now consider  an infinitesimal sudden quench $\lambda\rightarrow\lambda+\delta$,
which changes $H(\lambda)$ into the new Hamiltonian $H(\lambda+\delta)$ whose new ground
state we denote by $\ket{\psi_{0}(\lambda+\delta)}$. The irreversible work is defined as the difference between the work produced by the quench and the free energy difference \cite{PhysRevLett.78.2690}
\begin{equation}\label{Eq.Wirr}
\left\langle W_{irr}\right\rangle =\left\langle W\right\rangle -\Delta F~.
\end{equation}
Notice that the irreversible work, quantifying the deviation from a reversible, isothermal process is, at $T=0$, equivalent both to what has been dubbed {\textit{inner friction}} in Ref.~\cite{Plastina2014}, which quantifies the deviation from a reversible, adiabatic process, and to the {\textit{excess work}}, as reported in ref~\cite{Silva2008}. This equivalence, which however holds only at zero temperature, allows to interpret the following results also as the energetic cost of the adiabatic deviation of a sudden quench.

In the case of a sudden quench Eq.~\ref{Eq.Wirr} reads
\begin{equation} \label{eqn:wirr}
\left\langle W_{irr}\right\rangle =\braket{\psi_{0}(\lambda)|H(\lambda+\delta)|\psi_{0}(\lambda)}-\braket{\psi_{0}(\lambda+\delta)|H(\lambda+\delta)|\psi_{0}(\lambda+\delta)}.
\end{equation}
It has been shown \cite{Mascarenhas2014} that for a small quench, the irreversible work is proportional to the second derivative of the ground-state energy
\begin{equation}\label{eqn:iwderivative}
\left\langle W_{irr}\right\rangle =-\frac{\delta^{2}}{2}\frac{\partial^{2}}{\partial\lambda^{2}}E_{0}(\lambda).
\end{equation}

In  order  to deal with a quantity independent of the quench amplitude, it can be useful to consider the 
RIW instead  \cite{Bayat2016}  $\tilde{W}_{irr}=\left\langle W_{irr}\right\rangle/\delta^2$.
In this section we derive an analytical expression for the RIW. As a first step, we  calculate the first derivative of the ground state energy.  By means of the Hellmann-Feymann theorem
and taking into account that 
\begin{equation}
\frac{\partial}{\partial\lambda}H(\lambda)=H_{1}, 
\end{equation}
we get 
\begin{equation}
\frac{\partial}{\partial\lambda}E_{0}(\lambda)=\braket{\psi_{0}(\lambda)|H_{1}|\psi_{0}(\lambda)},
\end{equation}
To calculate the second derivative we can exploit the normalization condition $\braket{\psi_{0}(\lambda)|\psi_{0}(\lambda)}=1$ corresponding to
\begin{equation}
\braket{\psi_{0}(\lambda)|\frac{\partial}{\partial\lambda}\psi_{0}(\lambda)}+\braket{\frac{\partial}{\partial\lambda}\psi_{0}(\lambda)|\psi_{0}(\lambda)}=0, 
\end{equation}
and, after  a lengthy but straightforward calculation, we obtain

\begin{equation}\label{eqn:d2E0}
\frac{\partial^{2}}{\partial\lambda^{2}}E_{0}(\lambda)=2E_{0}(\lambda)\left[\braket{\frac{\partial}{\partial\lambda}\psi_{0}(\lambda)|\frac{\partial}{\partial\lambda}\psi_{0}(\lambda)}-\left|\braket{\frac{\partial}{\partial\lambda}\psi_{0}(\lambda)|\psi_{0}(\lambda)}\right|^{2}\right]-2\sum_{n\neq0}E_{n}(\lambda)\left|\left\langle \psi_{n}(\lambda)|\frac{\partial}{\partial\lambda}\psi_{0}(\lambda)|\right\rangle \right|^{2}.
\end{equation}
Comparing (\ref{eqn:d2E0}) with (\ref{eq:chiF}) and (\ref{eqn:pertchi}), we arrive to the main result of this section
\begin{equation}\label{eqn:comp}
\average{\tilde{W}_{irr}}=-\left[E_{0}(\lambda)\chi_{F}-\sum_{n\neq0}E_{n}(\lambda)\left|\frac{\braket{\psi_{n}(\lambda)|H_{1}|\psi_{0}(\lambda)}}{E_{0}(\lambda)-E_{n}(\lambda)}\right|^{2}\right].
\end{equation}
Equation (\ref{eqn:comp}) gives an explicit functional dependence of the RIW from the FS. The two quantities result to be not simply proportional because of the presence of the second term where the transitions to all the excited levels are taken into account.  It is worth noticing that in the second term of (\ref{eqn:comp}), the excited energies appear as a weight factor to the second order correction terms. A class of interesting  models would be the ones where the $\left\{  E_n(\lambda) \right\}$  almost coincide, also motivated by the recent interest for systems with very narrow energy band \cite{Huber2010,Jo2012,Gori2014,Valiente2015}. In this case one would expect that     $\average{\tilde{W}_{irr}} \propto  \chi_{F}$.
Such limit case could be the subject of future investigations.

We stress that Eq. (\ref{eqn:comp})  is only valid for sudden quenches, when Eq. (\ref{eqn:wirr}) holds.
In this kind of processes, we expect for the irreversible work to be always positive. 
Vanishing irreversible work can be obtained by sudden quenches only if $\left[H_1,H_0\right]=0 $, but in this case, in absence of level crossing, one should have $\chi_F=W_{irr}=0$  as well as $ \braket{\psi_{n}(\lambda)|H_{1}|\psi_{0}(\lambda)}=0$ for $n\neq0$. This would lead to no contradiction in Eq. (\ref{eqn:wirr}). 
In the next section we want to verify that this difference between RIW and FS gives a contribution also in a many-body system and in the thermodynamical limit.

\section{Irreversible work and fidelity susceptibility near quantum critical point in a quantum Ising chain }\label{S.Irr}

It has been shown that both the $\chi_F$ and the $\left\langle \tilde{W}_{irr} \right\rangle$ are able to signal a second order quantum phase transition   \cite{Dorner2012, Damskibook,DamskiRamsJPA14}, nevertheless they are expected to behave differently close to the quantum critical point. The scaling of $\chi_F$ near a critical point has been studied in \cite{Gu2008a,Gu2011} .
In this section we focus on a quantum Ising chain studying the scaling with the number of sites of the RIW   and comparing it with the scaling of the $\chi_F$. 

The Quantum Ising model in 1D consists on a N-site lattice of localized spin-$1/2$ particles, with nearest-neighbors Ising-like interaction along a given direction (say $x$)  and locally coupled to an external field along a perpendicular direction (say $z$) \cite{sachdev}
\begin{equation}\label{eqn:Qising}
H=-\sum_{j=1}^{N}\left[\lambda\sigma_{j}^{z}+\sigma_{j}^{x}\sigma_{j+1}^{x}\right].
\end{equation}
We impose periodic boundary conditions so that $\vec{\sigma}_1=\vec{\sigma}_{N+1}$. 
The Hamiltonian (\ref{eqn:Qising}) can be diagonalized exactly  by standard procedures. The first step consists in  applying the Jordan-Wigner transformation  (JWT)
mapping the spins into strings of spinless fermions. Denoting with $c_j$ and $c^\dagger_j$ the fermionic creation and annihilation operators, the JWT operates as follow 
\begin{eqnarray}
c_{j} & = & \prod_{l<j}\left(\sigma_{z}^{j}\right)\sigma_{+}^{j},\nonumber \\
c_{j}^{\dagger} & = & \prod_{l<j}\left(\sigma_{z}^{j}\right)\sigma_{-}^{j},\nonumber  \\
c^\dagger_{j}c_{j} & = & \frac{1-\sigma_{z}^{j}}{2},
\end{eqnarray}
The so-obtained Hamiltonian does not preserves the total number of spinless fermions but commutes with the parity operator. So the Hilbert space can be decomposed into
orthogonal parity subspaces. Hereafter, we will consider only the even-parity subspace, since it does not affect the results in the thermodynamical limit and, for every finite even $N$, it is also the exact ground state~\cite{DamskiRamsJPA14}. All the details of the diagonalization procedure are well known (see for example \cite{LIEB1961407}) and we will go only through the main points. 

After applying the JWT, we take advantage of the translational invariance of the model and 
perform a Fourier transformation
\begin{equation}
c_{j}=\frac{1}{\sqrt{N}}\sum_{k}e^{-ikj} d_{k}.
\end{equation}
In the even-parity sector,  the wavenumber is  $k\in K^{+}$, with $K^{\pm}=\left\{ k=\pm\frac{\pi\left(2n-1\right)}{N};\:n=1,\ldots,\frac{N}{2}\right\}$.
Finally, a Bogoliubov transformation
\begin{equation}
d_{\pm k}=\gamma_{\pm k} \cos \frac{\phi_k}{2}\mp \gamma^\dagger_{\mp k} \sin \frac{\phi_k}{2},
\end{equation}
with
\begin{equation}\label{eqn:angles}
 \cos \phi_k  =  2 \frac{\lambda-\cos k}{\epsilon_k},~
 \sin \phi_k  =  2\frac{\sin k}{\epsilon_k},
\end{equation}
diagonalizes  the Hamiltonian which, in the new representation, becomes
\begin{equation}
H=\sum_{k\in K^{+}} \epsilon_k \left(\gamma^\dagger_k \gamma_k - \frac{1}{2}\right).
\end{equation}
The quasi-particle spectrum is given by
\begin{equation}
\epsilon_{k}(\lambda)=2\sqrt{1+\lambda^{2}-2\lambda\cos k},
\end{equation}
and 
the ground state and its energy are given respectively by
\begin{equation}\label{eqn:gs}
\ket{gs}=\prod_{k>0}\left(\cos\frac{\phi_{k}}{2}+id_{k}^{\dagger}d_{-k}^{\dagger}\sin\frac{\phi_{k}}{2}\right)\ket{0}.
\end{equation}
and
\begin{equation}\label{eqn:gsen}
E_{0}(\lambda)=-\frac{1}{2}\sum_{k\in K^{+}}\epsilon_{k}=-\sum_{k\in K^{+}}\sqrt{1+\lambda^{2}-2\lambda\cos k}\simeq-\frac{N}{\pi}\int_{\frac{\pi}{N}}^{\frac{\pi(N-1)}{N}}dk\sqrt{1+\lambda^{2}-2\lambda\cos k}.
\end{equation}

Starting from these expressions, we will study the properties of the system at $T=0$. We want to investigate the behavior of the RIW near the CP and compare it with that of the FS, whose properties at criticality for the Ising model has been already investigated (see, e.g., Ref.~\cite{Gu2010, Damskibook} and references therein). 
As for the FS, the RIW is expected to have an extensive regular part scaling with $N$. It is also known \cite{SilvaPRL08, Dorner2012} that the RIW diverges close to the critical point but, differently from the FS, we will show that its divergence is logarithmic in $\lambda$.  As a consequence, their finite-size scaling behavior is not equivalent.

\subsection{Fidelity susceptibility near the critical point}

We start reviewing the scaling analysis that has been done for the $\chi_F$  in Ref.~\cite{Gu2010}.   From Eqs.(\ref{eq:chiF}) and (\ref{eqn:gs}) it is possible to see that  
\begin{equation}
\chi_{F}=\frac{1}{4}\sum_{k>0}\left(\frac{d\phi_{k}}{d\lambda}\right)^{2},
\end{equation}
and, using Eqs.~\ref{eqn:angles}, \ref{eqn:gsen}, \ref{eqn:gs} one obtains
\begin{equation}\label{eqn:errechi}
\chi_{F}=\frac{1}{4}\sum_{k>0}\frac{\sin^{2}k}{\left(1+\lambda^{2}-2\lambda\cos k\right)^{2}}\simeq\frac{N}{4\pi}\int_{\frac{\pi}{N}}^{\frac{\pi(N-1)}{N}}dk\frac{\sin^{2}k}{\left(1+\lambda^{2}-2\lambda\cos k\right)^{2}}=\frac{N}{4\pi} R_F(\lambda,N).
\end{equation}
Here, the integral $R_F(\lambda,N)$ represents the continuum  limit obtained by fixing the lattice number and increasing $N$, with $dk\simeq \frac{2\pi}{N}$.

It has been shown that in the limit of large $N$,  for    $\lambda<1$
\begin{equation}
\chi_{F}\simeq N\frac{1}{16\left(1-\lambda^{2}\right)},
\end{equation}
whereas for $\lambda>1$
\begin{equation}
\chi_{F}\simeq N\frac{1}{16\lambda^{2}\left(\lambda^{2}-1\right)},
\end{equation}
meaning that, close to the critical point, the quantity $\chi_{F}/N$ is intensive with a critical exponent $\alpha=1$.

For a finite system, in correspondence to the maximum at $\lambda=1$ the $\chi_F$ diverges faster
\begin{equation}
\chi_{F}\propto N^{2},
\end{equation}
with a critical adiabatic dimension $d_c=2$.

Exact analytical expressions for $\chi_F$ has been given in \cite{DamskiRamsJPA14}.

\subsection{Irreversible work}

\begin{figure}[h]
\centering
\includegraphics[width=3.3 cm,angle=-90]{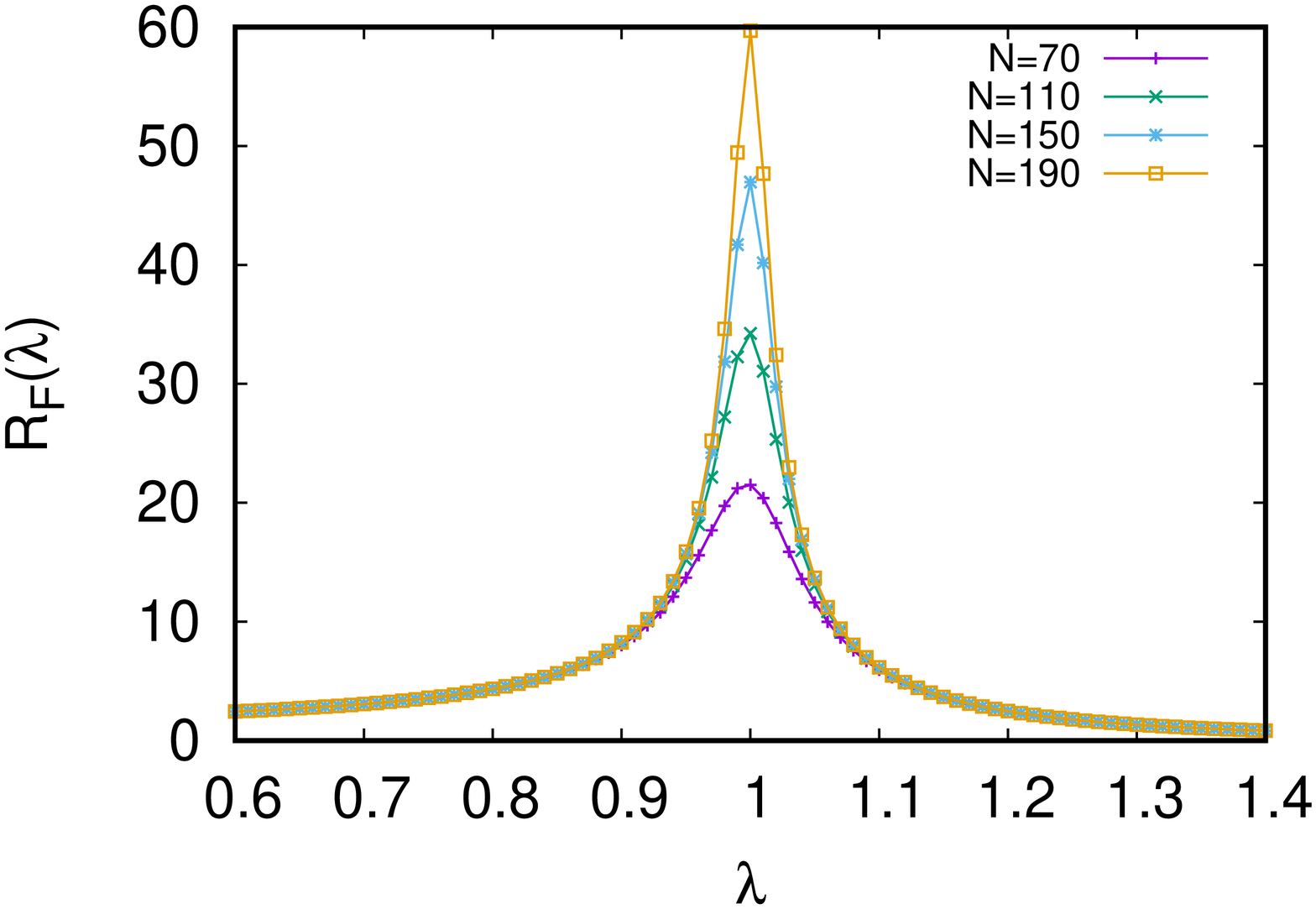}
\includegraphics[width=3.3 cm,angle=-90]{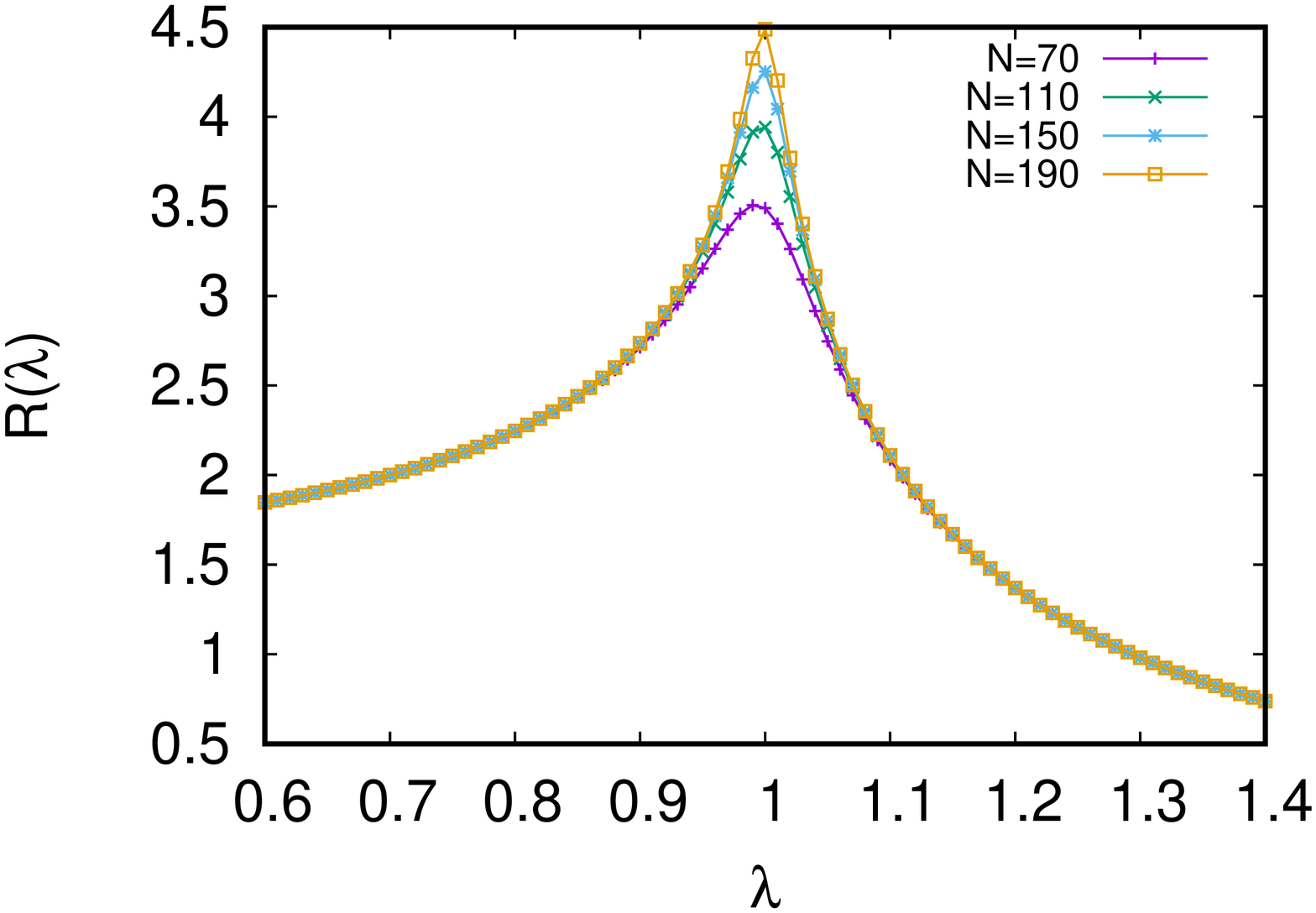}
\includegraphics[width=3.3 cm,angle=-90]{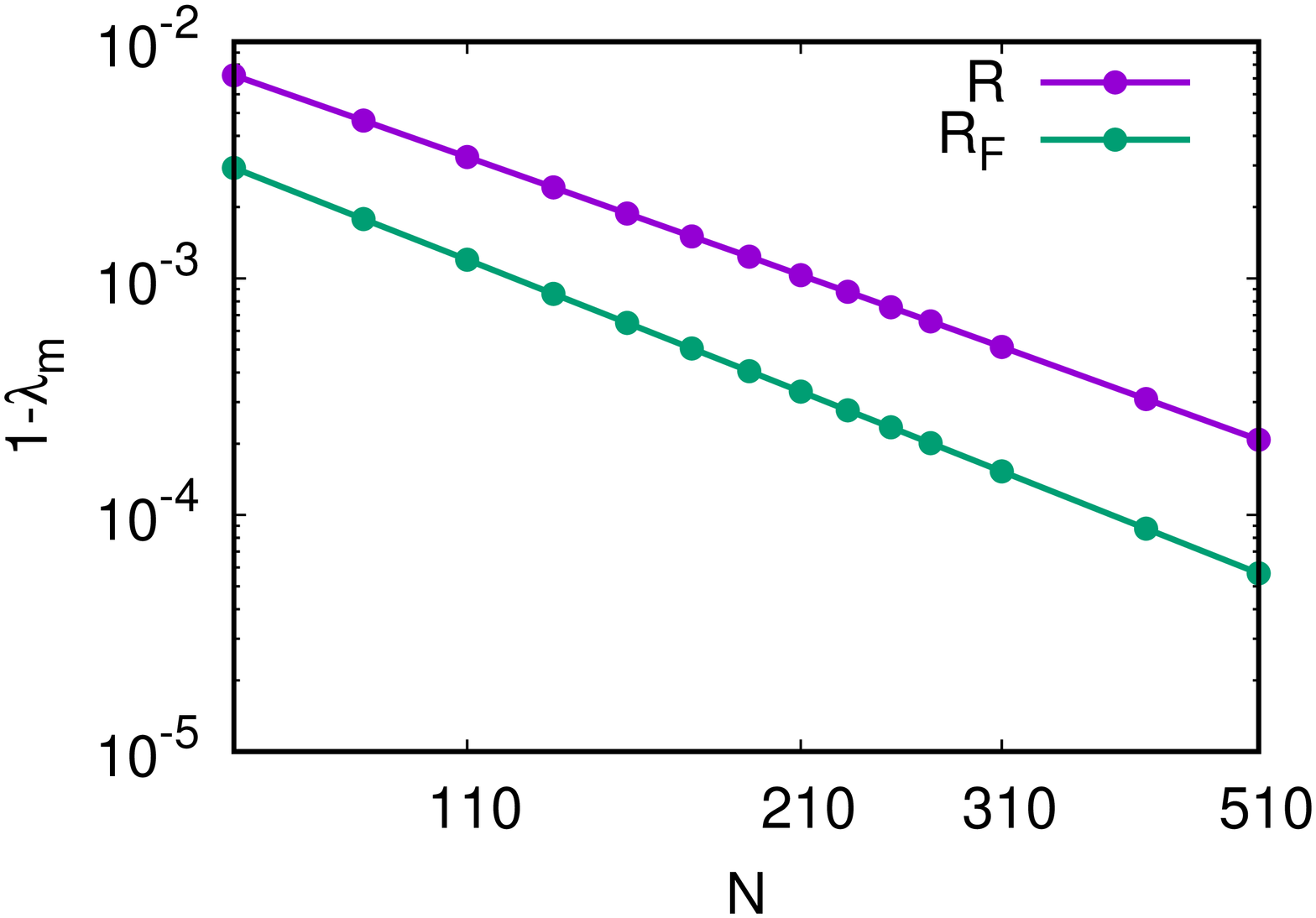}
\caption{Comparison between $ \chi_F 4\pi / N $ (left panel)  and  $ \left\langle \tilde{W}_{irr} \right\rangle  2 \pi /  N$  (central panel) close to the critical point, for different system sizes. It is evident how the maximum value at the finite-size critical point $\lambda_m(N)$ for the FS increases faster that that of the RIW. On the right panel we report in a log-log plot the difference between the infinite and the finite-size critical point $1-\lambda_m(N)$, as detected by FS and RIW respectively, as a function of $N$, showing that the location of the peak of $ \chi_F 4\pi / N $ approaches faster the critical point by increasing the system size.}
\label{fig:cfr} 
\end{figure}

In this last subsection we want to show that the RIW behaves differently from the FS, also in the thermodynamical limit where the second term in Eq. \ref{eqn:comp} still contributes. We will perform a similar analysis as in the previos subsection. 
Starting from  Eq. (\ref{eqn:iwderivative}) and using  (\ref{eqn:gsen}) we get

\begin{equation}
\frac{d^{2}E_{0}}{d\lambda^{2}}=-8\sum_{k\in K^{+}}\frac{\sin^{2}k}{\epsilon_{k}^{3}},
\end{equation}
and, in the continuum limit,
\begin{equation}
\average{\tilde{W}_{irr}}=4 \sum_{k\in K^{+}}\frac{\sin^{2}k}{\epsilon_{k}^{3}}
\simeq \frac{N}{2\pi}R(\lambda,N), 
\end{equation}
where we have defined the quantity
\begin{equation}\label{eqn:erre}
R(\lambda,N)=\int_{\frac{\pi}{N}}^{\frac{\pi(N-1)}{N}}dk\, \frac{\sin^{2}k}{\left(1+\lambda^{2}-2\lambda\cos k\right)^{\frac{3}{2}}},
\end{equation}
which, for the quantum Ising model, is proportional to the transverse magnetic susceptibility \cite{Um2007a,Gambassi2011}. 
In Fig. \ref{fig:cfr} a comparison between the two quantities $R_F(\lambda,n)$ and $R(\lambda,n)$  is reported. One can see that  both quantities are extensive away from the critical point, but close to the CP they develop divergences which grow differently with $N$. By comparing Eq.~\ref{eqn:erre} with Eq.~\ref{eqn:errechi}, we can conclude that this is due to the different exponents in the respective denominators which yields a logarithmic and an algebraic divergence, respectively, induced by the low energy modes $k\rightarrow 0$ for $\lambda=1$. 
Indeed, the function $R(\lambda,\infty)$ diverges only at $\lambda=1$ where
\begin{equation}
R(1,N)=\frac{1}{2^{\frac{3}{2}}}\int_{\frac{\pi}{N}}^{\frac{\pi(N-1)}{N}}dk\frac{\sin^{2}k}{\left(1-\cos k\right)^{\frac{3}{2}}}\simeq\frac{1}{2^{\frac{1}{2}}}\int_{\frac{\pi}{N}}^{\frac{\pi(N-1)}{N}}dk\frac{1}{k}\propto\ln N.
\end{equation}
Therefore, for a finite size-system,  the irreversible work evaluated at its maximum,  grows logarithmically with the system's size
\begin{equation}
\average{\tilde{W}_{irr}}\propto N \ln N,
\end{equation}
corresponding to critical adiabatic dimension equal to zero.

It is well known~\cite{Barber1983} that the peaks in the vicinity of a CP for finite size systems are due to the presence of a QPT in the thermodynamical limit and the relative scaling with the system size is related to the nature of the divergence at criticality for an infinite system. In the following we will show that, in the thermodynamical limit, the RIW exhibits indeed a logarithmic divergence by letting $\lambda$ approach the CP of the Ising QPT. This is a direct consequence of the above-stated proportionality between the RIW and the transverse magnetic susceptibility \cite{Gambassi2011} for the quantum Ising model. The latter then can be mapped to a 2D classical Ising model where the specific heat at constant volume corresponds to the transverse magnetic susceptibility of the quantum model.

For $\lambda\neq 1$ 
 it is useful to express the energy as
\begin{equation}\label{eqn:gselliptic}
E_{0}(\lambda)  \simeq  -\frac{2N}{\pi}(1+\lambda)
\mathcal{E}\left(\frac{4\lambda}{(1+\lambda)^{2}}\right)
\end{equation}
where the function $\mathcal{E}$ is the complete elliptic integral of the second 
kind \cite{Abramowitz} 
\begin{equation}
\mathcal{E}(m)=\mathcal{E}\left(\frac{\pi}{2}|m\right)\label{eq:ellipt2compl},
\end{equation}
defined from the incomplete elliptic integral of the second kind 
\begin{equation}
\mathcal{E}\left(\phi|m\right)=\int_{0}^{\phi}d\theta\left(1-m\sin^{2}\theta\right)^{\frac{1}{2}},
\end{equation}
with $0\leq m\leq1$.
The first two derivatives  of $\mathcal{E}(m)$ are respectively
\begin{equation}
\frac{\partial \mathcal{E}(m)}{\partial m}=\frac{\mathcal{E}(m)-\mathcal{K}(m)}{2m},
\end{equation}
\begin{equation}
\frac{\partial^{2}\mathcal{E}(m)}{\partial m^{2}}=\frac{2(m-1)\mathcal{K}(m)-(m-2)\mathcal{E}(m)}{4(m-1)m^{2}},
\end{equation}
where $\mathcal{K}$ is the complete elliptic integral of the first kind
\begin{equation}
\mathcal{K}(m)=\mathcal{K}\left(\frac{\pi}{2}|m\right),
\end{equation}
defined from the incomplete elliptic integral of the first kind
\begin{equation}
\mathcal{K}(\phi|m)=\int_{0}^{\phi}\frac{d\theta}{\sqrt{1-m\sin^{2}\theta}}, 
\end{equation}
with $0\leq m\leq1$.

The RIW is obtained from the second derivative of (\ref{eqn:gselliptic}) whose expression in terms of elliptic functions is 
\begin{equation}
\left\langle \tilde{W}_{irr}\right\rangle   \simeq -\frac{N}{\pi}\frac{\left(1+\lambda^{2}\right)\mathcal{K}\left(\frac{4\lambda}{(1+\lambda)^{2}}\right)-\left(1+\lambda\right)^{2}\mathcal{E}\left(\frac{4\lambda}{(1+\lambda)^{2}}\right)}{2\lambda^{2}(1+\lambda)}.
\end{equation}
We notice that $m=\frac{4\lambda}{(1+\lambda)^{2}}\leq1$ so  for $\lambda\rightarrow1$ one has
$m\rightarrow1^{-}$. Using the asymptotic limits  
$\mathcal{E}(1)=1$ and $\mathcal{K}(m\rightarrow1^{-})\sim-\frac{\ln\left(1-m\right)}{2}$
we finally obtain 
\begin{equation}
\left\langle \tilde{W}_{irr}\right\rangle   \simeq  -\frac{N}{2\pi}\mathcal{K}(1^{-})\simeq
  \frac{N}{4\pi}\ln\left(1-\frac{4\lambda}{(1+\lambda)^{2}}\right)\propto N \ln\left|1-\lambda\right|.
\end{equation}
As expected, the RIW  behaves extensively but $\left\langle \tilde{W}_{irr}\right\rangle$ has a critical exponent $\alpha=0$ and  diverges logarithmically at $\lambda=1$.

\begin{figure}[h]
\centering
\includegraphics[width=5 cm,angle=-90]{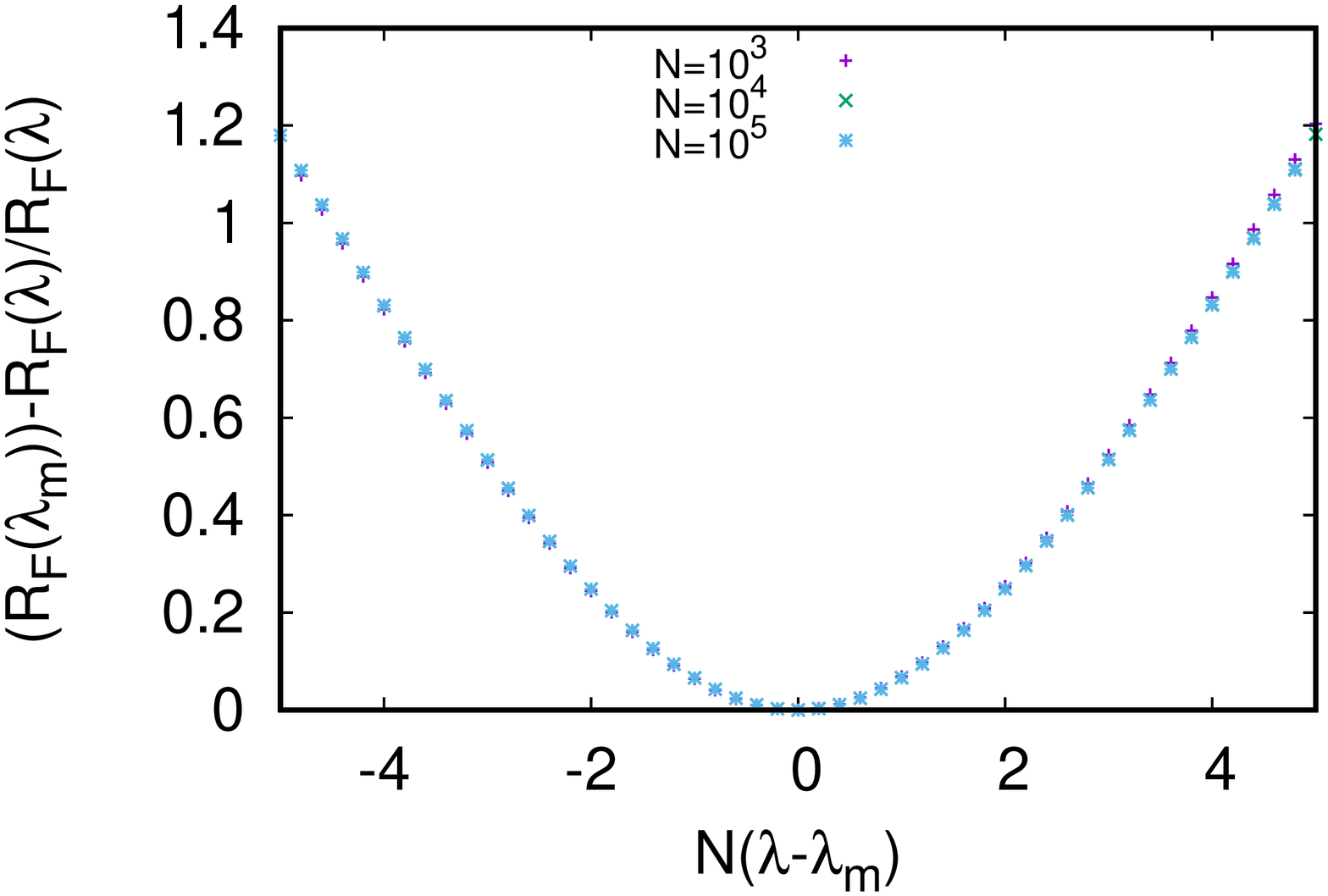}
\includegraphics[width=5 cm,angle=-90]{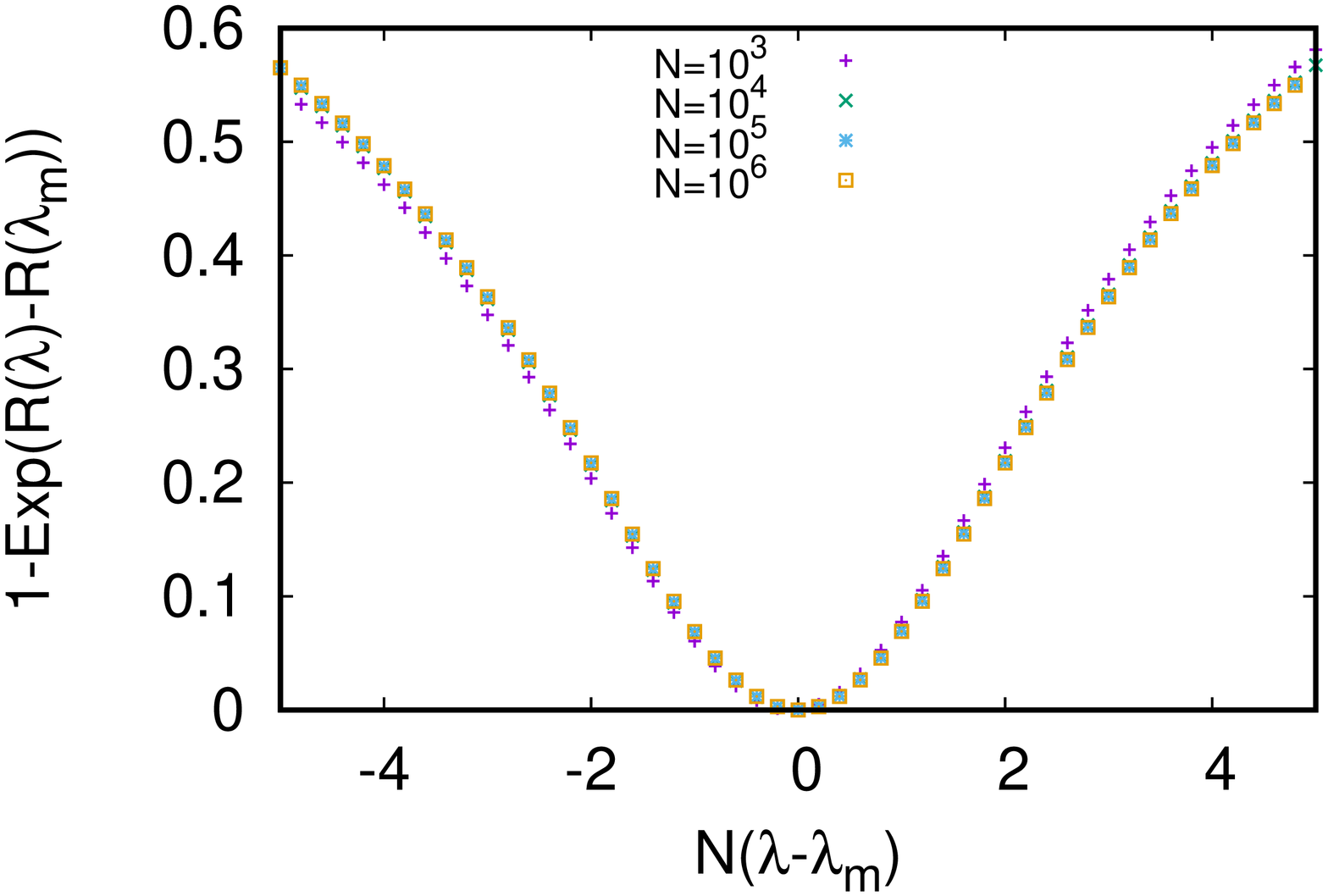}
\caption{Data collapse of the FS and the RIW according to Eqs.~\ref{E.collF} and \ref{E.collW} on the right and left panel respectively. The scaling function  of the factor $R_F(\lambda)$ and $R(\lambda)$ are considered as a function of  the parameter $N^{\frac{1}{\nu}}(\lambda-\lambda_m)$  with critical exponent $\nu = 1$ in both cases. Notice, however, the different finite-size scaling Ans{\"a}tze on the $y$-axis for the algebraic and the logarithmic case.}
\label{fig:datcoll}
\end{figure}

Therefore, it is possible to perform a finite size scaling for the case of logarithmic divergences with the scaling function  \cite{Barber1983}  
\begin{equation}\label{E.collW}
1-\exp\left[ R(\lambda,N)-R(\lambda_m,N)\right]=f(N^{\frac{1}{\nu}}(\lambda-\lambda_m)), 
\end{equation}
being $\nu=1$ the critical exponent of the correlation length for the transverse Ising model. The data collapse is depicted in Fig. \ref{fig:datcoll} together with the finite-size scaling analysis of the fidelity susceptibility exhibiting the algebraic divergence, where the finite-size scaling Ansatz reads
\begin{equation}\label{E.collF}
\frac{R_F(\lambda,N)-R_f(\lambda_m,N)}{R_f(\lambda_m,N)}=g(N^{\frac{1}{\nu}}(\lambda-\lambda_m)).
\end{equation}

A further step to the present analysis would be the study of the finite-temperature case \cite{Dorner,Gu2010, Sotiriadis2013}. To see again the functional difference between RIW and FS, one has to generalize Eq.  (\ref{eqn:comp}) to the $T \neq 0$ case. It would be also interesting to study the behavior of these two quantities 
in ancillary systems introduced as a probe \cite{giorgi2010}, so to have experimental access to the phase transition. 

\section{Conclusions}\label{S.Concl}
Given a quantum system with a tunable parameter, we related the irreversible work produced by an infinitesimal quench at zero temperature to the fidelity susceptibility of the system's ground state.  We showed that the irreversible work differs from the fidelity susceptibility by an extra term which takes into account the transitions to excited states of the pre-quenched system. 
We compared the behavior of these two quantities close to a second-order quantum phase transition point where they are known to diverge.
In this regard, we considered a 1D quantum Ising model and analytically studied how the irreversible work scales in the thermodynamical limit comparing it with the fidelity susceptibility. We have seen that both have an extensive  regular part, but, while the fidelity susceptibility has a critical exponent $\alpha=1$ and a critical adiabatic dimension $d_c=2$, the irreversible work diverges logarithmically at the critical point, meaning a critical exponent $\alpha=0$. The presence of an algebraic and a logarithmic divergence at criticality for the fidelity susceptibility and the irreversible work, respectively, has been traced back to different exponents in the respective perturbation series expansion. Finally, relying on the finite-size scaling theory, we performed data collapse for both quantities evaluated at different system sizes. From this analysis, we can conclude that the fidelity susceptibility allows for a better detection of the critical point in finite-size systems as both its maximum value and the location of the critical point obey scaling relations that converge faster to the infinite-size case.

To conclude, it would be interesting to investigate if similar results can be obtained also for higher order quantum phase transitions where the critical point may be detected at higher orders in the perturbation expansion of the ground state fidelity.

\section{Acknowledgments}
We acknowledge fruitful discussions with Gabriele De Chiara and Abolfazl Bayat. SP is supported by a Rita Levi-Montalcini fellowship of MIUR and T.J.G.A. acknowledges funding  under the EU Collaborative Project TherMiQ (Grant No. 618074) and the UE grant  QuPRoCs (Grant Agreement 641277).

\bibliographystyle{apsrev4-1} 
\bibliography{qtherm} 

\begin{thebibliography}{51}%
\makeatletter
\providecommand \@ifxundefined [1]{%
 \@ifx{#1\undefined}
}%
\providecommand \@ifnum [1]{%
 \ifnum #1\expandafter \@firstoftwo
 \else \expandafter \@secondoftwo
 \fi
}%
\providecommand \@ifx [1]{%
 \ifx #1\expandafter \@firstoftwo
 \else \expandafter \@secondoftwo
 \fi
}%
\providecommand \natexlab [1]{#1}%
\providecommand \enquote  [1]{``#1''}%
\providecommand \bibnamefont  [1]{#1}%
\providecommand \bibfnamefont [1]{#1}%
\providecommand \citenamefont [1]{#1}%
\providecommand \href@noop [0]{\@secondoftwo}%
\providecommand \href [0]{\begingroup \@sanitize@url \@href}%
\providecommand \@href[1]{\@@startlink{#1}\@@href}%
\providecommand \@@href[1]{\endgroup#1\@@endlink}%
\providecommand \@sanitize@url [0]{\catcode `\\12\catcode `\$12\catcode
  `\&12\catcode `\#12\catcode `\^12\catcode `\_12\catcode `\%12\relax}%
\providecommand \@@startlink[1]{}%
\providecommand \@@endlink[0]{}%
\providecommand \url  [0]{\begingroup\@sanitize@url \@url }%
\providecommand \@url [1]{\endgroup\@href {#1}{\urlprefix }}%
\providecommand \urlprefix  [0]{URL }%
\providecommand \Eprint [0]{\href }%
\providecommand \doibase [0]{http://dx.doi.org/}%
\providecommand \selectlanguage [0]{\@gobble}%
\providecommand \bibinfo  [0]{\@secondoftwo}%
\providecommand \bibfield  [0]{\@secondoftwo}%
\providecommand \translation [1]{[#1]}%
\providecommand \BibitemOpen [0]{}%
\providecommand \bibitemStop [0]{}%
\providecommand \bibitemNoStop [0]{.\EOS\space}%
\providecommand \EOS [0]{\spacefactor3000\relax}%
\providecommand \BibitemShut  [1]{\csname bibitem#1\endcsname}%
\let\auto@bib@innerbib\@empty
\bibitem [{\citenamefont {Goold}\ \emph {et~al.}(2016)\citenamefont {Goold},
  \citenamefont {Huber}, \citenamefont {Riera}, \citenamefont {del Rio},\ and\
  \citenamefont {Skrzypczyk}}]{Goold2016}%
  \BibitemOpen
  \bibfield  {author} {\bibinfo {author} {\bibfnamefont {J.}~\bibnamefont
  {Goold}}, \bibinfo {author} {\bibfnamefont {M.}~\bibnamefont {Huber}},
  \bibinfo {author} {\bibfnamefont {A.}~\bibnamefont {Riera}}, \bibinfo
  {author} {\bibfnamefont {L.}~\bibnamefont {del Rio}}, \ and\ \bibinfo
  {author} {\bibfnamefont {P.}~\bibnamefont {Skrzypczyk}},\ }\href
  {http://stacks.iop.org/1751-8121/49/i=14/a=143001} {\bibfield  {journal}
  {\bibinfo  {journal} {Journal of Physics A: Mathematical and Theoretical}\
  }\textbf {\bibinfo {volume} {49}},\ \bibinfo {pages} {143001} (\bibinfo
  {year} {2016})}\BibitemShut {NoStop}%
\bibitem [{\citenamefont {Kosloff}(2013)}]{Kosloff2016}%
  \BibitemOpen
  \bibfield  {author} {\bibinfo {author} {\bibfnamefont {R.}~\bibnamefont
  {Kosloff}},\ }\href {\doibase 10.3390/e15062100} {\bibfield  {journal}
  {\bibinfo  {journal} {Entropy}\ }\textbf {\bibinfo {volume} {15}},\ \bibinfo
  {pages} {2100} (\bibinfo {year} {2013})}\BibitemShut {NoStop}%
\bibitem [{\citenamefont {{Cheneau}}\ \emph {et~al.}(2012)\citenamefont
  {{Cheneau}}, \citenamefont {{Barmettler}}, \citenamefont {{Poletti}},
  \citenamefont {{Endres}}, \citenamefont {{Schau{\ss}}}, \citenamefont
  {{Fukuhara}}, \citenamefont {{Gross}}, \citenamefont {{Bloch}}, \citenamefont
  {{Kollath}},\ and\ \citenamefont {{Kuhr}}}]{BlochNat12}%
  \BibitemOpen
  \bibfield  {author} {\bibinfo {author} {\bibfnamefont {M.}~\bibnamefont
  {{Cheneau}}}, \bibinfo {author} {\bibfnamefont {P.}~\bibnamefont
  {{Barmettler}}}, \bibinfo {author} {\bibfnamefont {D.}~\bibnamefont
  {{Poletti}}}, \bibinfo {author} {\bibfnamefont {M.}~\bibnamefont {{Endres}}},
  \bibinfo {author} {\bibfnamefont {P.}~\bibnamefont {{Schau{\ss}}}}, \bibinfo
  {author} {\bibfnamefont {T.}~\bibnamefont {{Fukuhara}}}, \bibinfo {author}
  {\bibfnamefont {C.}~\bibnamefont {{Gross}}}, \bibinfo {author} {\bibfnamefont
  {I.}~\bibnamefont {{Bloch}}}, \bibinfo {author} {\bibfnamefont
  {C.}~\bibnamefont {{Kollath}}}, \ and\ \bibinfo {author} {\bibfnamefont
  {S.}~\bibnamefont {{Kuhr}}},\ }\href {\doibase 10.1038/nature10748}
  {\bibfield  {journal} {\bibinfo  {journal} {\nat}\ }\textbf {\bibinfo
  {volume} {481}},\ \bibinfo {pages} {484} (\bibinfo {year}
  {2012})}\BibitemShut {NoStop}%
\bibitem [{\citenamefont {Polkovnikov}\ \emph {et~al.}(2011)\citenamefont
  {Polkovnikov}, \citenamefont {Sengupta}, \citenamefont {Silva},\ and\
  \citenamefont {Vengalattore}}]{Polkovnikov2011}%
  \BibitemOpen
  \bibfield  {author} {\bibinfo {author} {\bibfnamefont {A.}~\bibnamefont
  {Polkovnikov}}, \bibinfo {author} {\bibfnamefont {K.}~\bibnamefont
  {Sengupta}}, \bibinfo {author} {\bibfnamefont {A.}~\bibnamefont {Silva}}, \
  and\ \bibinfo {author} {\bibfnamefont {M.}~\bibnamefont {Vengalattore}},\
  }\href {\doibase 10.1103/RevModPhys.83.863} {\bibfield  {journal} {\bibinfo
  {journal} {Rev. Mod. Phys.}\ }\textbf {\bibinfo {volume} {83}},\ \bibinfo
  {pages} {863} (\bibinfo {year} {2011})}\BibitemShut {NoStop}%
\bibitem [{\citenamefont {Jarzynski}(1997{\natexlab{a}})}]{Jarzynski1997}%
  \BibitemOpen
  \bibfield  {author} {\bibinfo {author} {\bibfnamefont {C.}~\bibnamefont
  {Jarzynski}},\ }\href {\doibase 10.1103/PhysRevLett.78.2690} {\bibfield
  {journal} {\bibinfo  {journal} {Phys. Rev. Lett.}\ }\textbf {\bibinfo
  {volume} {78}},\ \bibinfo {pages} {2690} (\bibinfo {year}
  {1997}{\natexlab{a}})}\BibitemShut {NoStop}%
\bibitem [{\citenamefont {Jarzynski}(2011)}]{Jarzynski2011}%
  \BibitemOpen
  \bibfield  {author} {\bibinfo {author} {\bibfnamefont {C.}~\bibnamefont
  {Jarzynski}},\ }\href@noop {} {\bibfield  {journal} {\bibinfo  {journal}
  {Annual Review of Condensed Matter Physics}\ }\textbf {\bibinfo {volume}
  {2}},\ \bibinfo {pages} {329} (\bibinfo {year} {2011})}\BibitemShut {NoStop}%
\bibitem [{\citenamefont {Campisi}\ \emph {et~al.}(2011)\citenamefont
  {Campisi}, \citenamefont {H{\"{a}}nggi},\ and\ \citenamefont
  {Talkner}}]{Campisi2011}%
  \BibitemOpen
  \bibfield  {author} {\bibinfo {author} {\bibfnamefont {M.}~\bibnamefont
  {Campisi}}, \bibinfo {author} {\bibfnamefont {P.}~\bibnamefont
  {H{\"{a}}nggi}}, \ and\ \bibinfo {author} {\bibfnamefont {P.}~\bibnamefont
  {Talkner}},\ }\href {\doibase 10.1103/RevModPhys.83.771} {\bibfield
  {journal} {\bibinfo  {journal} {Rev. Mod. Phys.}\ }\textbf {\bibinfo {volume}
  {83}},\ \bibinfo {pages} {771} (\bibinfo {year} {2011})}\BibitemShut
  {NoStop}%
\bibitem [{\citenamefont {Toyabe}\ \emph {et~al.}(2010)\citenamefont {Toyabe},
  \citenamefont {Sagawa}, \citenamefont {Ueda}, \citenamefont {Muneyuki},\ and\
  \citenamefont {Sano}}]{Toyabe2010}%
  \BibitemOpen
  \bibfield  {author} {\bibinfo {author} {\bibfnamefont {S.}~\bibnamefont
  {Toyabe}}, \bibinfo {author} {\bibfnamefont {T.}~\bibnamefont {Sagawa}},
  \bibinfo {author} {\bibfnamefont {M.}~\bibnamefont {Ueda}}, \bibinfo {author}
  {\bibfnamefont {E.}~\bibnamefont {Muneyuki}}, \ and\ \bibinfo {author}
  {\bibfnamefont {M.}~\bibnamefont {Sano}},\ }\href@noop {} {\bibfield
  {journal} {\bibinfo  {journal} {Nat. Phys.}\ }\textbf {\bibinfo {volume}
  {7}},\ \bibinfo {pages} {988} (\bibinfo {year} {2010})}\BibitemShut {NoStop}%
\bibitem [{\citenamefont {An}\ \emph {et~al.}(2014)\citenamefont {An},
  \citenamefont {Zhang}, \citenamefont {Um}, \citenamefont {Lv}, \citenamefont
  {Lu}, \citenamefont {Zhang}, \citenamefont {Yin}, \citenamefont {Quan},\ and\
  \citenamefont {Kim}}]{An2014}%
  \BibitemOpen
  \bibfield  {author} {\bibinfo {author} {\bibfnamefont {S.}~\bibnamefont
  {An}}, \bibinfo {author} {\bibfnamefont {J.-N.}\ \bibnamefont {Zhang}},
  \bibinfo {author} {\bibfnamefont {M.}~\bibnamefont {Um}}, \bibinfo {author}
  {\bibfnamefont {D.}~\bibnamefont {Lv}}, \bibinfo {author} {\bibfnamefont
  {Y.}~\bibnamefont {Lu}}, \bibinfo {author} {\bibfnamefont {J.}~\bibnamefont
  {Zhang}}, \bibinfo {author} {\bibfnamefont {Z.-Q.}\ \bibnamefont {Yin}},
  \bibinfo {author} {\bibfnamefont {H.~T.}\ \bibnamefont {Quan}}, \ and\
  \bibinfo {author} {\bibfnamefont {K.}~\bibnamefont {Kim}},\ }\href@noop {}
  {\bibfield  {journal} {\bibinfo  {journal} {Nat. Phys.}\ }\textbf {\bibinfo
  {volume} {11}},\ \bibinfo {pages} {193} (\bibinfo {year} {2014})}\BibitemShut
  {NoStop}%
\bibitem [{\citenamefont {{Brunelli}}\ \emph {et~al.}(2016)\citenamefont
  {{Brunelli}}, \citenamefont {{Fusco}}, \citenamefont {{Landig}},
  \citenamefont {{Wieczorek}}, \citenamefont {{Hoelscher-Obermaier}},
  \citenamefont {{Landi}}, \citenamefont {{Semiao}}, \citenamefont {{Ferraro}},
  \citenamefont {{Kiesel}}, \citenamefont {{Donner}}, \citenamefont {{De
  Chiara}},\ and\ \citenamefont {{Paternostro}}}]{2016arXiv160206958B}%
  \BibitemOpen
  \bibfield  {author} {\bibinfo {author} {\bibfnamefont {M.}~\bibnamefont
  {{Brunelli}}}, \bibinfo {author} {\bibfnamefont {L.}~\bibnamefont {{Fusco}}},
  \bibinfo {author} {\bibfnamefont {R.}~\bibnamefont {{Landig}}}, \bibinfo
  {author} {\bibfnamefont {W.}~\bibnamefont {{Wieczorek}}}, \bibinfo {author}
  {\bibfnamefont {J.}~\bibnamefont {{Hoelscher-Obermaier}}}, \bibinfo {author}
  {\bibfnamefont {G.}~\bibnamefont {{Landi}}}, \bibinfo {author} {\bibfnamefont
  {F.~L.}\ \bibnamefont {{Semiao}}}, \bibinfo {author} {\bibfnamefont
  {A.}~\bibnamefont {{Ferraro}}}, \bibinfo {author} {\bibfnamefont
  {N.}~\bibnamefont {{Kiesel}}}, \bibinfo {author} {\bibfnamefont
  {T.}~\bibnamefont {{Donner}}}, \bibinfo {author} {\bibfnamefont
  {G.}~\bibnamefont {{De Chiara}}}, \ and\ \bibinfo {author} {\bibfnamefont
  {M.}~\bibnamefont {{Paternostro}}},\ }\href@noop {} {\bibfield  {journal}
  {\bibinfo  {journal} {ArXiv e-prints}\ } (\bibinfo {year} {2016})},\ \Eprint
  {http://arxiv.org/abs/1602.06958} {arXiv:1602.06958 [quant-ph]} \BibitemShut
  {NoStop}%
\bibitem [{\citenamefont {Batalh\~ao}\ \emph {et~al.}(2014)\citenamefont
  {Batalh\~ao}, \citenamefont {Souza}, \citenamefont {Mazzola}, \citenamefont
  {Auccaise}, \citenamefont {Sarthour}, \citenamefont {Oliveira}, \citenamefont
  {Goold}, \citenamefont {De~Chiara}, \citenamefont {Paternostro},\ and\
  \citenamefont {Serra}}]{PhysRevLett.113.140601}%
  \BibitemOpen
  \bibfield  {author} {\bibinfo {author} {\bibfnamefont {T.~B.}\ \bibnamefont
  {Batalh\~ao}}, \bibinfo {author} {\bibfnamefont {A.~M.}\ \bibnamefont
  {Souza}}, \bibinfo {author} {\bibfnamefont {L.}~\bibnamefont {Mazzola}},
  \bibinfo {author} {\bibfnamefont {R.}~\bibnamefont {Auccaise}}, \bibinfo
  {author} {\bibfnamefont {R.~S.}\ \bibnamefont {Sarthour}}, \bibinfo {author}
  {\bibfnamefont {I.~S.}\ \bibnamefont {Oliveira}}, \bibinfo {author}
  {\bibfnamefont {J.}~\bibnamefont {Goold}}, \bibinfo {author} {\bibfnamefont
  {G.}~\bibnamefont {De~Chiara}}, \bibinfo {author} {\bibfnamefont
  {M.}~\bibnamefont {Paternostro}}, \ and\ \bibinfo {author} {\bibfnamefont
  {R.~M.}\ \bibnamefont {Serra}},\ }\href {\doibase
  10.1103/PhysRevLett.113.140601} {\bibfield  {journal} {\bibinfo  {journal}
  {Phys. Rev. Lett.}\ }\textbf {\bibinfo {volume} {113}},\ \bibinfo {pages}
  {140601} (\bibinfo {year} {2014})}\BibitemShut {NoStop}%
\bibitem [{\citenamefont {Dorner}\ \emph {et~al.}(2013)\citenamefont {Dorner},
  \citenamefont {Clark}, \citenamefont {Heaney}, \citenamefont {Fazio},
  \citenamefont {Goold},\ and\ \citenamefont
  {Vedral}}]{PhysRevLett.110.230601}%
  \BibitemOpen
  \bibfield  {author} {\bibinfo {author} {\bibfnamefont {R.}~\bibnamefont
  {Dorner}}, \bibinfo {author} {\bibfnamefont {S.~R.}\ \bibnamefont {Clark}},
  \bibinfo {author} {\bibfnamefont {L.}~\bibnamefont {Heaney}}, \bibinfo
  {author} {\bibfnamefont {R.}~\bibnamefont {Fazio}}, \bibinfo {author}
  {\bibfnamefont {J.}~\bibnamefont {Goold}}, \ and\ \bibinfo {author}
  {\bibfnamefont {V.}~\bibnamefont {Vedral}},\ }\href {\doibase
  10.1103/PhysRevLett.110.230601} {\bibfield  {journal} {\bibinfo  {journal}
  {Phys. Rev. Lett.}\ }\textbf {\bibinfo {volume} {110}},\ \bibinfo {pages}
  {230601} (\bibinfo {year} {2013})}\BibitemShut {NoStop}%
\bibitem [{\citenamefont {Batalh\~ao}\ \emph {et~al.}(2015)\citenamefont
  {Batalh\~ao}, \citenamefont {Souza}, \citenamefont {Sarthour}, \citenamefont
  {Oliveira}, \citenamefont {Paternostro}, \citenamefont {Lutz},\ and\
  \citenamefont {Serra}}]{PhysRevLett.115.190601}%
  \BibitemOpen
  \bibfield  {author} {\bibinfo {author} {\bibfnamefont {T.~B.}\ \bibnamefont
  {Batalh\~ao}}, \bibinfo {author} {\bibfnamefont {A.~M.}\ \bibnamefont
  {Souza}}, \bibinfo {author} {\bibfnamefont {R.~S.}\ \bibnamefont {Sarthour}},
  \bibinfo {author} {\bibfnamefont {I.~S.}\ \bibnamefont {Oliveira}}, \bibinfo
  {author} {\bibfnamefont {M.}~\bibnamefont {Paternostro}}, \bibinfo {author}
  {\bibfnamefont {E.}~\bibnamefont {Lutz}}, \ and\ \bibinfo {author}
  {\bibfnamefont {R.~M.}\ \bibnamefont {Serra}},\ }\href {\doibase
  10.1103/PhysRevLett.115.190601} {\bibfield  {journal} {\bibinfo  {journal}
  {Phys. Rev. Lett.}\ }\textbf {\bibinfo {volume} {115}},\ \bibinfo {pages}
  {190601} (\bibinfo {year} {2015})}\BibitemShut {NoStop}%
\bibitem [{\citenamefont {{Gambassi}}\ and\ \citenamefont
  {{Silva}}(2011)}]{Gambassi2011}%
  \BibitemOpen
  \bibfield  {author} {\bibinfo {author} {\bibfnamefont {A.}~\bibnamefont
  {{Gambassi}}}\ and\ \bibinfo {author} {\bibfnamefont {A.}~\bibnamefont
  {{Silva}}},\ }\href@noop {} {\bibfield  {journal} {\bibinfo  {journal} {ArXiv
  e-prints}\ } (\bibinfo {year} {2011})},\ \Eprint
  {http://arxiv.org/abs/1106.2671} {arXiv:1106.2671} \BibitemShut {NoStop}%
\bibitem [{\citenamefont {Plastina}\ \emph {et~al.}(2014)\citenamefont
  {Plastina}, \citenamefont {Alecce}, \citenamefont {Apollaro}, \citenamefont
  {Falcone}, \citenamefont {Francica}, \citenamefont {Galve}, \citenamefont
  {Gullo},\ and\ \citenamefont {Zambrini}}]{Plastina2014}%
  \BibitemOpen
  \bibfield  {author} {\bibinfo {author} {\bibfnamefont {F.}~\bibnamefont
  {Plastina}}, \bibinfo {author} {\bibfnamefont {a.}~\bibnamefont {Alecce}},
  \bibinfo {author} {\bibfnamefont {T.~J.~G.}\ \bibnamefont {Apollaro}},
  \bibinfo {author} {\bibfnamefont {G.}~\bibnamefont {Falcone}}, \bibinfo
  {author} {\bibfnamefont {G.}~\bibnamefont {Francica}}, \bibinfo {author}
  {\bibfnamefont {F.}~\bibnamefont {Galve}}, \bibinfo {author} {\bibfnamefont
  {N.~L.}\ \bibnamefont {Gullo}}, \ and\ \bibinfo {author} {\bibfnamefont
  {R.}~\bibnamefont {Zambrini}},\ }\href {\doibase
  10.1103/PhysRevLett.113.260601} {\bibfield  {journal} {\bibinfo  {journal}
  {Phys. Rev. Lett.}\ }\textbf {\bibinfo {volume} {113}},\ \bibinfo {pages}
  {260601} (\bibinfo {year} {2014})}\BibitemShut {NoStop}%
\bibitem [{\citenamefont {Fusco}\ \emph {et~al.}(2014)\citenamefont {Fusco},
  \citenamefont {Pigeon}, \citenamefont {Apollaro}, \citenamefont {Xuereb},
  \citenamefont {Mazzola}, \citenamefont {Campisi}, \citenamefont {Ferraro},
  \citenamefont {Paternostro},\ and\ \citenamefont {{De Chiara}}}]{Fusco2014}%
  \BibitemOpen
  \bibfield  {author} {\bibinfo {author} {\bibfnamefont {L.}~\bibnamefont
  {Fusco}}, \bibinfo {author} {\bibfnamefont {S.}~\bibnamefont {Pigeon}},
  \bibinfo {author} {\bibfnamefont {T.~J.~G.}\ \bibnamefont {Apollaro}},
  \bibinfo {author} {\bibfnamefont {A.}~\bibnamefont {Xuereb}}, \bibinfo
  {author} {\bibfnamefont {L.}~\bibnamefont {Mazzola}}, \bibinfo {author}
  {\bibfnamefont {M.}~\bibnamefont {Campisi}}, \bibinfo {author} {\bibfnamefont
  {A.}~\bibnamefont {Ferraro}}, \bibinfo {author} {\bibfnamefont
  {M.}~\bibnamefont {Paternostro}}, \ and\ \bibinfo {author} {\bibfnamefont
  {G.}~\bibnamefont {{De Chiara}}},\ }\href {\doibase
  10.1103/PhysRevX.4.031029} {\bibfield  {journal} {\bibinfo  {journal} {Phys.
  Rev. X}\ }\textbf {\bibinfo {volume} {4}},\ \bibinfo {pages} {13} (\bibinfo
  {year} {2014})}\BibitemShut {NoStop}%
\bibitem [{\citenamefont {Talkner}\ \emph {et~al.}(2007)\citenamefont
  {Talkner}, \citenamefont {Lutz},\ and\ \citenamefont
  {H{\"{a}}nggi}}]{Talkner2007a}%
  \BibitemOpen
  \bibfield  {author} {\bibinfo {author} {\bibfnamefont {P.}~\bibnamefont
  {Talkner}}, \bibinfo {author} {\bibfnamefont {E.}~\bibnamefont {Lutz}}, \
  and\ \bibinfo {author} {\bibfnamefont {P.}~\bibnamefont {H{\"{a}}nggi}},\
  }\href {\doibase 10.1103/PhysRevE.75.050102} {\bibfield  {journal} {\bibinfo
  {journal} {Phys. Rev. E}\ }\textbf {\bibinfo {volume} {75}},\ \bibinfo
  {pages} {050102} (\bibinfo {year} {2007})}\BibitemShut {NoStop}%
\bibitem [{\citenamefont {Silva}(2008{\natexlab{a}})}]{Silva2008}%
  \BibitemOpen
  \bibfield  {author} {\bibinfo {author} {\bibfnamefont {A.}~\bibnamefont
  {Silva}},\ }\href {\doibase 10.1103/PhysRevLett.101.120603} {\bibfield
  {journal} {\bibinfo  {journal} {Phys. Rev. Lett.}\ }\textbf {\bibinfo
  {volume} {101}},\ \bibinfo {pages} {120603} (\bibinfo {year}
  {2008}{\natexlab{a}})}\BibitemShut {NoStop}%
\bibitem [{\citenamefont {Dorner}\ \emph {et~al.}(2012)\citenamefont {Dorner},
  \citenamefont {Goold}, \citenamefont {Cormick}, \citenamefont {Paternostro},\
  and\ \citenamefont {Vedral}}]{Dorner2012}%
  \BibitemOpen
  \bibfield  {author} {\bibinfo {author} {\bibfnamefont {R.}~\bibnamefont
  {Dorner}}, \bibinfo {author} {\bibfnamefont {J.}~\bibnamefont {Goold}},
  \bibinfo {author} {\bibfnamefont {C.}~\bibnamefont {Cormick}}, \bibinfo
  {author} {\bibfnamefont {M.}~\bibnamefont {Paternostro}}, \ and\ \bibinfo
  {author} {\bibfnamefont {V.}~\bibnamefont {Vedral}},\ }\href {\doibase
  10.1103/PhysRevLett.109.160601} {\bibfield  {journal} {\bibinfo  {journal}
  {Phys. Rev. Lett.}\ }\textbf {\bibinfo {volume} {109}},\ \bibinfo {pages}
  {160601} (\bibinfo {year} {2012})}\BibitemShut {NoStop}%
\bibitem [{\citenamefont {Gambassi}\ and\ \citenamefont
  {Silva}(2012)}]{Gambassi2012}%
  \BibitemOpen
  \bibfield  {author} {\bibinfo {author} {\bibfnamefont {A.}~\bibnamefont
  {Gambassi}}\ and\ \bibinfo {author} {\bibfnamefont {A.}~\bibnamefont
  {Silva}},\ }\href@noop {} {\bibfield  {journal} {\bibinfo  {journal} {Phys.
  Rev. Lett.}\ }\textbf {\bibinfo {volume} {109}},\ \bibinfo {pages} {250602}
  (\bibinfo {year} {2012})}\BibitemShut {NoStop}%
\bibitem [{\citenamefont {Apollaro}\ \emph {et~al.}(2015)\citenamefont
  {Apollaro}, \citenamefont {Francica}, \citenamefont {Paternostro},\ and\
  \citenamefont {Campisi}}]{Apollaro2015}%
  \BibitemOpen
  \bibfield  {author} {\bibinfo {author} {\bibfnamefont {T.~J.~G.}\
  \bibnamefont {Apollaro}}, \bibinfo {author} {\bibfnamefont {G.}~\bibnamefont
  {Francica}}, \bibinfo {author} {\bibfnamefont {M.}~\bibnamefont
  {Paternostro}}, \ and\ \bibinfo {author} {\bibfnamefont {M.}~\bibnamefont
  {Campisi}},\ }\href@noop {} {\bibfield  {journal} {\bibinfo  {journal} {Phys.
  Scr.}\ }\textbf {\bibinfo {volume} {T165}},\ \bibinfo {pages} {014023}
  (\bibinfo {year} {2015})}\BibitemShut {NoStop}%
\bibitem [{\citenamefont {Sindona}\ \emph {et~al.}(2014)\citenamefont
  {Sindona}, \citenamefont {Goold}, \citenamefont {Lo~Gullo},\ and\
  \citenamefont {Plastina}}]{Sindona2014}%
  \BibitemOpen
  \bibfield  {author} {\bibinfo {author} {\bibfnamefont {A.}~\bibnamefont
  {Sindona}}, \bibinfo {author} {\bibfnamefont {J.}~\bibnamefont {Goold}},
  \bibinfo {author} {\bibfnamefont {N.}~\bibnamefont {Lo~Gullo}}, \ and\
  \bibinfo {author} {\bibfnamefont {F.}~\bibnamefont {Plastina}},\ }\href
  {http://stacks.iop.org/1367-2630/16/i=4/a=045013} {\bibfield  {journal}
  {\bibinfo  {journal} {New Journal of Physics}\ }\textbf {\bibinfo {volume}
  {16}},\ \bibinfo {pages} {045013} (\bibinfo {year} {2014})}\BibitemShut
  {NoStop}%
\bibitem [{\citenamefont {Apollaro}\ \emph {et~al.}(2016)\citenamefont
  {Apollaro}, \citenamefont {Palma},\ and\ \citenamefont
  {Marino}}]{Apollaro2016}%
  \BibitemOpen
  \bibfield  {author} {\bibinfo {author} {\bibfnamefont {T.~J.~G.}\
  \bibnamefont {Apollaro}}, \bibinfo {author} {\bibfnamefont {G.~M.}\
  \bibnamefont {Palma}}, \ and\ \bibinfo {author} {\bibfnamefont
  {J.}~\bibnamefont {Marino}},\ }\href@noop {} {\  (\bibinfo {year} {2016})},\
  \Eprint {http://arxiv.org/abs/1603.03579} {arXiv:1603.03579} \BibitemShut
  {NoStop}%
\bibitem [{\citenamefont {Mascarenhas}\ \emph {et~al.}(2014)\citenamefont
  {Mascarenhas}, \citenamefont {Bragan\ifmmode~\mbox{\c{c}}\else \c{c}\fi{}a},
  \citenamefont {Dorner}, \citenamefont {Fran\ifmmode
  \mbox{\c{c}}\else~\c{c}\fi{}a Santos}, \citenamefont {Vedral}, \citenamefont
  {Modi},\ and\ \citenamefont {Goold}}]{Mascarenhas2014}%
  \BibitemOpen
  \bibfield  {author} {\bibinfo {author} {\bibfnamefont {E.}~\bibnamefont
  {Mascarenhas}}, \bibinfo {author} {\bibfnamefont {H.}~\bibnamefont
  {Bragan\ifmmode~\mbox{\c{c}}\else \c{c}\fi{}a}}, \bibinfo {author}
  {\bibfnamefont {R.}~\bibnamefont {Dorner}}, \bibinfo {author} {\bibfnamefont
  {M.}~\bibnamefont {Fran\ifmmode \mbox{\c{c}}\else~\c{c}\fi{}a Santos}},
  \bibinfo {author} {\bibfnamefont {V.}~\bibnamefont {Vedral}}, \bibinfo
  {author} {\bibfnamefont {K.}~\bibnamefont {Modi}}, \ and\ \bibinfo {author}
  {\bibfnamefont {J.}~\bibnamefont {Goold}},\ }\href {\doibase
  10.1103/PhysRevE.89.062103} {\bibfield  {journal} {\bibinfo  {journal} {Phys.
  Rev. E}\ }\textbf {\bibinfo {volume} {89}},\ \bibinfo {pages} {062103}
  (\bibinfo {year} {2014})}\BibitemShut {NoStop}%
\bibitem [{\citenamefont {{Bayat}}\ \emph {et~al.}(2016)\citenamefont
  {{Bayat}}, \citenamefont {{Apollaro}}, \citenamefont {{Paganelli}},
  \citenamefont {{De Chiara}}, \citenamefont {{Johannesson}}, \citenamefont
  {{Bose}},\ and\ \citenamefont {{Sodano}}}]{Bayat2016}%
  \BibitemOpen
  \bibfield  {author} {\bibinfo {author} {\bibfnamefont {A.}~\bibnamefont
  {{Bayat}}}, \bibinfo {author} {\bibfnamefont {T.~J.~G.}\ \bibnamefont
  {{Apollaro}}}, \bibinfo {author} {\bibfnamefont {S.}~\bibnamefont
  {{Paganelli}}}, \bibinfo {author} {\bibfnamefont {G.}~\bibnamefont {{De
  Chiara}}}, \bibinfo {author} {\bibfnamefont {H.}~\bibnamefont
  {{Johannesson}}}, \bibinfo {author} {\bibfnamefont {S.}~\bibnamefont
  {{Bose}}}, \ and\ \bibinfo {author} {\bibfnamefont {P.}~\bibnamefont
  {{Sodano}}},\ }\href@noop {} {\bibfield  {journal} {\bibinfo  {journal}
  {Phys. Rev. B}\ }\textbf {\bibinfo {volume} {93}},\ \bibinfo {pages} {201106}
  (\bibinfo {year} {2016})}\BibitemShut {NoStop}%
\bibitem [{\citenamefont {Quan}\ \emph {et~al.}(2006)\citenamefont {Quan},
  \citenamefont {Song}, \citenamefont {Liu}, \citenamefont {Zanardi},\ and\
  \citenamefont {Sun}}]{Quan2006}%
  \BibitemOpen
  \bibfield  {author} {\bibinfo {author} {\bibfnamefont {H.~T.}\ \bibnamefont
  {Quan}}, \bibinfo {author} {\bibfnamefont {Z.}~\bibnamefont {Song}}, \bibinfo
  {author} {\bibfnamefont {X.~F.}\ \bibnamefont {Liu}}, \bibinfo {author}
  {\bibfnamefont {P.}~\bibnamefont {Zanardi}}, \ and\ \bibinfo {author}
  {\bibfnamefont {C.~P.}\ \bibnamefont {Sun}},\ }\href {\doibase
  10.1103/PhysRevLett.96.140604} {\bibfield  {journal} {\bibinfo  {journal}
  {Phys. Rev. Lett.}\ }\textbf {\bibinfo {volume} {96}},\ \bibinfo {pages}
  {140604} (\bibinfo {year} {2006})}\BibitemShut {NoStop}%
\bibitem [{\citenamefont {Zanardi}\ and\ \citenamefont
  {Paunkovi\ifmmode~\acute{c}\else
  \'{c}\fi{}}(2006{\natexlab{a}})}]{Zanardi2006}%
  \BibitemOpen
  \bibfield  {author} {\bibinfo {author} {\bibfnamefont {P.}~\bibnamefont
  {Zanardi}}\ and\ \bibinfo {author} {\bibfnamefont {N.}~\bibnamefont
  {Paunkovi\ifmmode~\acute{c}\else \'{c}\fi{}}},\ }\href {\doibase
  10.1103/PhysRevE.74.031123} {\bibfield  {journal} {\bibinfo  {journal} {Phys.
  Rev. E}\ }\textbf {\bibinfo {volume} {74}},\ \bibinfo {pages} {031123}
  (\bibinfo {year} {2006}{\natexlab{a}})}\BibitemShut {NoStop}%
\bibitem [{\citenamefont {You}\ \emph {et~al.}(2007)\citenamefont {You},
  \citenamefont {Li},\ and\ \citenamefont {Gu}}]{You2007}%
  \BibitemOpen
  \bibfield  {author} {\bibinfo {author} {\bibfnamefont {W.-L.}\ \bibnamefont
  {You}}, \bibinfo {author} {\bibfnamefont {Y.-W.}\ \bibnamefont {Li}}, \ and\
  \bibinfo {author} {\bibfnamefont {S.-J.}\ \bibnamefont {Gu}},\ }\href
  {\doibase 10.1103/PhysRevE.76.022101} {\bibfield  {journal} {\bibinfo
  {journal} {Phys. Rev. E}\ }\textbf {\bibinfo {volume} {76}},\ \bibinfo
  {pages} {022101} (\bibinfo {year} {2007})}\BibitemShut {NoStop}%
\bibitem [{\citenamefont {Campos~Venuti}\ and\ \citenamefont
  {Zanardi}(2007)}]{Campos2007}%
  \BibitemOpen
  \bibfield  {author} {\bibinfo {author} {\bibfnamefont {L.}~\bibnamefont
  {Campos~Venuti}}\ and\ \bibinfo {author} {\bibfnamefont {P.}~\bibnamefont
  {Zanardi}},\ }\href {\doibase 10.1103/PhysRevLett.99.095701} {\bibfield
  {journal} {\bibinfo  {journal} {Phys. Rev. Lett.}\ }\textbf {\bibinfo
  {volume} {99}},\ \bibinfo {pages} {095701} (\bibinfo {year}
  {2007})}\BibitemShut {NoStop}%
\bibitem [{\citenamefont {Gu}(2010)}]{Gu2010}%
  \BibitemOpen
  \bibfield  {author} {\bibinfo {author} {\bibfnamefont {S.-J.}\ \bibnamefont
  {Gu}},\ }\href@noop {} {\bibfield  {journal} {\bibinfo  {journal} {Int. J.
  Mod. Phys. B}\ }\textbf {\bibinfo {volume} {24}},\ \bibinfo {pages} {4371}
  (\bibinfo {year} {2010})}\BibitemShut {NoStop}%
\bibitem [{\citenamefont {Gu}\ \emph {et~al.}(2008)\citenamefont {Gu},
  \citenamefont {Kwok}, \citenamefont {Ning},\ and\ \citenamefont
  {Lin}}]{Gu2008a}%
  \BibitemOpen
  \bibfield  {author} {\bibinfo {author} {\bibfnamefont {S.-J.}\ \bibnamefont
  {Gu}}, \bibinfo {author} {\bibfnamefont {H.-M.}\ \bibnamefont {Kwok}},
  \bibinfo {author} {\bibfnamefont {W.-Q.}\ \bibnamefont {Ning}}, \ and\
  \bibinfo {author} {\bibfnamefont {H.-Q.}\ \bibnamefont {Lin}},\ }\href
  {\doibase 10.1103/PhysRevB.77.245109} {\bibfield  {journal} {\bibinfo
  {journal} {Physical Review B}\ }\textbf {\bibinfo {volume} {77}},\ \bibinfo
  {pages} {245109} (\bibinfo {year} {2008})}\BibitemShut {NoStop}%
\bibitem [{\citenamefont {Gu}\ \emph {et~al.}(2011)\citenamefont {Gu},
  \citenamefont {Kwok}, \citenamefont {Ning},\ and\ \citenamefont
  {Lin}}]{Gu2011}%
  \BibitemOpen
  \bibfield  {author} {\bibinfo {author} {\bibfnamefont {S.-J.}\ \bibnamefont
  {Gu}}, \bibinfo {author} {\bibfnamefont {H.-M.}\ \bibnamefont {Kwok}},
  \bibinfo {author} {\bibfnamefont {W.-Q.}\ \bibnamefont {Ning}}, \ and\
  \bibinfo {author} {\bibfnamefont {H.-Q.}\ \bibnamefont {Lin}},\ }\href
  {\doibase 10.1103/PhysRevB.83.159905} {\bibfield  {journal} {\bibinfo
  {journal} {Physical Review B}\ }\textbf {\bibinfo {volume} {83}},\ \bibinfo
  {pages} {159905} (\bibinfo {year} {2011})}\BibitemShut {NoStop}%
\bibitem [{\citenamefont {Uhlmann}(2011)}]{Uhlmann2011}%
  \BibitemOpen
  \bibfield  {author} {\bibinfo {author} {\bibfnamefont {A.}~\bibnamefont
  {Uhlmann}},\ }\href {\doibase 10.1007/s10701-009-9381-y} {\bibfield
  {journal} {\bibinfo  {journal} {Foundations of Physics}\ }\textbf {\bibinfo
  {volume} {41}},\ \bibinfo {pages} {288} (\bibinfo {year} {2011})}\BibitemShut
  {NoStop}%
\bibitem [{\citenamefont {Zanardi}\ and\ \citenamefont
  {Paunkovi\ifmmode~\acute{c}\else
  \'{c}\fi{}}(2006{\natexlab{b}})}]{ZanardiPRE06}%
  \BibitemOpen
  \bibfield  {author} {\bibinfo {author} {\bibfnamefont {P.}~\bibnamefont
  {Zanardi}}\ and\ \bibinfo {author} {\bibfnamefont {N.}~\bibnamefont
  {Paunkovi\ifmmode~\acute{c}\else \'{c}\fi{}}},\ }\href {\doibase
  10.1103/PhysRevE.74.031123} {\bibfield  {journal} {\bibinfo  {journal} {Phys.
  Rev. E}\ }\textbf {\bibinfo {volume} {74}},\ \bibinfo {pages} {031123}
  (\bibinfo {year} {2006}{\natexlab{b}})}\BibitemShut {NoStop}%
\bibitem [{\citenamefont {Zanardi}\ \emph {et~al.}(2007)\citenamefont
  {Zanardi}, \citenamefont {Giorda},\ and\ \citenamefont
  {Cozzini}}]{PhysRevLett.99.100603}%
  \BibitemOpen
  \bibfield  {author} {\bibinfo {author} {\bibfnamefont {P.}~\bibnamefont
  {Zanardi}}, \bibinfo {author} {\bibfnamefont {P.}~\bibnamefont {Giorda}}, \
  and\ \bibinfo {author} {\bibfnamefont {M.}~\bibnamefont {Cozzini}},\ }\href
  {\doibase 10.1103/PhysRevLett.99.100603} {\bibfield  {journal} {\bibinfo
  {journal} {Phys. Rev. Lett.}\ }\textbf {\bibinfo {volume} {99}},\ \bibinfo
  {pages} {100603} (\bibinfo {year} {2007})}\BibitemShut {NoStop}%
\bibitem [{\citenamefont
  {Jarzynski}(1997{\natexlab{b}})}]{PhysRevLett.78.2690}%
  \BibitemOpen
  \bibfield  {author} {\bibinfo {author} {\bibfnamefont {C.}~\bibnamefont
  {Jarzynski}},\ }\href {\doibase 10.1103/PhysRevLett.78.2690} {\bibfield
  {journal} {\bibinfo  {journal} {Phys. Rev. Lett.}\ }\textbf {\bibinfo
  {volume} {78}},\ \bibinfo {pages} {2690} (\bibinfo {year}
  {1997}{\natexlab{b}})}\BibitemShut {NoStop}%
\bibitem [{\citenamefont {Huber}\ and\ \citenamefont
  {Altman}(2010)}]{Huber2010}%
  \BibitemOpen
  \bibfield  {author} {\bibinfo {author} {\bibfnamefont {S.~D.}\ \bibnamefont
  {Huber}}\ and\ \bibinfo {author} {\bibfnamefont {E.}~\bibnamefont {Altman}},\
  }\href {\doibase 10.1103/PhysRevB.82.184502} {\bibfield  {journal} {\bibinfo
  {journal} {Phys. Rev. B}\ }\textbf {\bibinfo {volume} {82}},\ \bibinfo
  {pages} {184502} (\bibinfo {year} {2010})}\BibitemShut {NoStop}%
\bibitem [{\citenamefont {Jo}\ \emph {et~al.}(2012)\citenamefont {Jo},
  \citenamefont {Guzman}, \citenamefont {Thomas}, \citenamefont {Hosur},
  \citenamefont {Vishwanath},\ and\ \citenamefont {Stamper-Kurn}}]{Jo2012}%
  \BibitemOpen
  \bibfield  {author} {\bibinfo {author} {\bibfnamefont {G.-B.}\ \bibnamefont
  {Jo}}, \bibinfo {author} {\bibfnamefont {J.}~\bibnamefont {Guzman}}, \bibinfo
  {author} {\bibfnamefont {C.~K.}\ \bibnamefont {Thomas}}, \bibinfo {author}
  {\bibfnamefont {P.}~\bibnamefont {Hosur}}, \bibinfo {author} {\bibfnamefont
  {A.}~\bibnamefont {Vishwanath}}, \ and\ \bibinfo {author} {\bibfnamefont
  {D.~M.}\ \bibnamefont {Stamper-Kurn}},\ }\href {\doibase
  10.1103/PhysRevLett.108.045305} {\bibfield  {journal} {\bibinfo  {journal}
  {Phys. Rev. Lett.}\ }\textbf {\bibinfo {volume} {108}},\ \bibinfo {pages}
  {045305} (\bibinfo {year} {2012})}\BibitemShut {NoStop}%
\bibitem [{\citenamefont {{Gori}}\ \emph {et~al.}(2015)\citenamefont {{Gori}},
  \citenamefont {{Paganelli}}, \citenamefont {{Sharma}}, \citenamefont
  {{Sodano}},\ and\ \citenamefont {{Trombettoni}}}]{Gori2014}%
  \BibitemOpen
  \bibfield  {author} {\bibinfo {author} {\bibfnamefont {G.}~\bibnamefont
  {{Gori}}}, \bibinfo {author} {\bibfnamefont {S.}~\bibnamefont {{Paganelli}}},
  \bibinfo {author} {\bibfnamefont {A.}~\bibnamefont {{Sharma}}}, \bibinfo
  {author} {\bibfnamefont {P.}~\bibnamefont {{Sodano}}}, \ and\ \bibinfo
  {author} {\bibfnamefont {A.}~\bibnamefont {{Trombettoni}}},\ }\href {\doibase
  10.1103/PhysRevB.91.245138} {\bibfield  {journal} {\bibinfo  {journal}
  {Physical Rev. B}\ }\textbf {\bibinfo {volume} {91}},\ \bibinfo {eid}
  {245138} (\bibinfo {year} {2015})}\BibitemShut {NoStop}%
\bibitem [{\citenamefont {{Phillips}}\ \emph {et~al.}(2015)\citenamefont
  {{Phillips}}, \citenamefont {{De Chiara}}, \citenamefont {{{\"O}hberg}},\
  and\ \citenamefont {{Valiente}}}]{Valiente2015}%
  \BibitemOpen
  \bibfield  {author} {\bibinfo {author} {\bibfnamefont {L.~G.}\ \bibnamefont
  {{Phillips}}}, \bibinfo {author} {\bibfnamefont {G.}~\bibnamefont {{De
  Chiara}}}, \bibinfo {author} {\bibfnamefont {P.}~\bibnamefont
  {{{\"O}hberg}}}, \ and\ \bibinfo {author} {\bibfnamefont {M.}~\bibnamefont
  {{Valiente}}},\ }\href {\doibase 10.1103/PhysRevB.91.054103} {\bibfield
  {journal} {\bibinfo  {journal} {Phys. Rev. B}\ }\textbf {\bibinfo {volume}
  {91}},\ \bibinfo {eid} {054103} (\bibinfo {year} {2015})}\BibitemShut
  {NoStop}%
\bibitem [{\citenamefont {Damski}(2015)}]{Damskibook}%
  \BibitemOpen
  \bibfield  {author} {\bibinfo {author} {\bibfnamefont {B.}~\bibnamefont
  {Damski}},\ }\enquote {\bibinfo {title} {Fidelity approach to quantum phase
  transitions in quantum ising model},}\ in\ \href {\doibase
  10.1142/9789814704090_0006} {\emph {\bibinfo {booktitle} {Quantum Criticality
  in Condensed Matter}}}\ (\bibinfo  {publisher} {WORLD SCIENTIFIC},\ \bibinfo
  {year} {2015})\ pp.\ \bibinfo {pages} {159--182}\BibitemShut {NoStop}%
\bibitem [{\citenamefont {Damski}\ and\ \citenamefont
  {Rams}(2014)}]{DamskiRamsJPA14}%
  \BibitemOpen
  \bibfield  {author} {\bibinfo {author} {\bibfnamefont {B.}~\bibnamefont
  {Damski}}\ and\ \bibinfo {author} {\bibfnamefont {M.~M.}\ \bibnamefont
  {Rams}},\ }\href {http://stacks.iop.org/1751-8121/47/i=2/a=025303} {\bibfield
   {journal} {\bibinfo  {journal} {Journal of Physics A: Mathematical and
  Theoretical}\ }\textbf {\bibinfo {volume} {47}},\ \bibinfo {pages} {025303}
  (\bibinfo {year} {2014})}\BibitemShut {NoStop}%
\bibitem [{\citenamefont {Sachdev}(1999)}]{sachdev}%
  \BibitemOpen
  \bibfield  {author} {\bibinfo {author} {\bibfnamefont {S.}~\bibnamefont
  {Sachdev}},\ }\href@noop {} {\emph {\bibinfo {title} {Quantum Phase
  Transitions}}}\ (\bibinfo  {publisher} {Cambridge University Press,},\
  \bibinfo {year} {1999})\BibitemShut {NoStop}%
\bibitem [{\citenamefont {Lieb}\ \emph {et~al.}(1961)\citenamefont {Lieb},
  \citenamefont {Schultz},\ and\ \citenamefont {Mattis}}]{LIEB1961407}%
  \BibitemOpen
  \bibfield  {author} {\bibinfo {author} {\bibfnamefont {E.}~\bibnamefont
  {Lieb}}, \bibinfo {author} {\bibfnamefont {T.}~\bibnamefont {Schultz}}, \
  and\ \bibinfo {author} {\bibfnamefont {D.}~\bibnamefont {Mattis}},\ }\href
  {\doibase http://dx.doi.org/10.1016/0003-4916(61)90115-4} {\bibfield
  {journal} {\bibinfo  {journal} {Annals of Physics}\ }\textbf {\bibinfo
  {volume} {16}},\ \bibinfo {pages} {407 } (\bibinfo {year}
  {1961})}\BibitemShut {NoStop}%
\bibitem [{\citenamefont {Silva}(2008{\natexlab{b}})}]{SilvaPRL08}%
  \BibitemOpen
  \bibfield  {author} {\bibinfo {author} {\bibfnamefont {A.}~\bibnamefont
  {Silva}},\ }\href {\doibase 10.1103/PhysRevLett.101.120603} {\bibfield
  {journal} {\bibinfo  {journal} {Phys. Rev. Lett.}\ }\textbf {\bibinfo
  {volume} {101}},\ \bibinfo {pages} {120603} (\bibinfo {year}
  {2008}{\natexlab{b}})}\BibitemShut {NoStop}%
\bibitem [{\citenamefont {Um}\ \emph {et~al.}(2007)\citenamefont {Um},
  \citenamefont {Lee},\ and\ \citenamefont {Kim}}]{Um2007a}%
  \BibitemOpen
  \bibfield  {author} {\bibinfo {author} {\bibfnamefont {J.}~\bibnamefont
  {Um}}, \bibinfo {author} {\bibfnamefont {S.-I.}\ \bibnamefont {Lee}}, \ and\
  \bibinfo {author} {\bibfnamefont {B.~J.}\ \bibnamefont {Kim}},\ }\href@noop
  {} {\bibfield  {journal} {\bibinfo  {journal} {Journal of the Korean Physical
  Society}\ }\textbf {\bibinfo {volume} {50}},\ \bibinfo {pages} {285}
  (\bibinfo {year} {2007})}\BibitemShut {NoStop}%
\bibitem [{\citenamefont {Barber}(1983)}]{Barber1983}%
  \BibitemOpen
  \bibfield  {author} {\bibinfo {author} {\bibfnamefont {M.~N.}\ \bibnamefont
  {Barber}},\ }\href@noop {} {\emph {\bibinfo {title} {Phase Transitions and
  Critical Phenomena}}},\ edited by\ \bibinfo {editor} {\bibfnamefont
  {C.}~\bibnamefont {Domb}}\ and\ \bibinfo {editor} {\bibfnamefont {J.~L.}\
  \bibnamefont {Leibovitz}},\ Vol.~\bibinfo {volume} {8}\ (\bibinfo
  {publisher} {cademic, London},\ \bibinfo {year} {1983})\ p.\ \bibinfo {pages}
  {146}\BibitemShut {NoStop}%
\bibitem [{\citenamefont {Abramowitz}\ and\ \citenamefont
  {Stegun}(1964)}]{Abramowitz}%
  \BibitemOpen
  \bibfield  {author} {\bibinfo {author} {\bibfnamefont {M.}~\bibnamefont
  {Abramowitz}}\ and\ \bibinfo {author} {\bibfnamefont {I.~A.}\ \bibnamefont
  {Stegun}},\ }\href@noop {} {\emph {\bibinfo {title} {Handbook of Mathematical
  Functions}}}\ (\bibinfo  {publisher} {Dover},\ \bibinfo {year}
  {1964})\BibitemShut {NoStop}%
\bibitem [{\citenamefont {Dorner}\ \emph {et~al.}()\citenamefont {Dorner},
  \citenamefont {Goold}, \citenamefont {Cormick}, \citenamefont {Paternostro},\
  and\ \citenamefont {Vedral}}]{Dorner}%
  \BibitemOpen
  \bibfield  {author} {\bibinfo {author} {\bibfnamefont {R.}~\bibnamefont
  {Dorner}}, \bibinfo {author} {\bibfnamefont {J.}~\bibnamefont {Goold}},
  \bibinfo {author} {\bibfnamefont {C.}~\bibnamefont {Cormick}}, \bibinfo
  {author} {\bibfnamefont {M.}~\bibnamefont {Paternostro}}, \ and\ \bibinfo
  {author} {\bibfnamefont {V.}~\bibnamefont {Vedral}},\ }\href@noop {} {\ ,\
  \bibinfo {pages} {3}}\BibitemShut {NoStop}%
\bibitem [{\citenamefont {Sotiriadis}\ \emph {et~al.}(2013)\citenamefont
  {Sotiriadis}, \citenamefont {Gambassi},\ and\ \citenamefont
  {Silva}}]{Sotiriadis2013}%
  \BibitemOpen
  \bibfield  {author} {\bibinfo {author} {\bibfnamefont {S.}~\bibnamefont
  {Sotiriadis}}, \bibinfo {author} {\bibfnamefont {A.}~\bibnamefont
  {Gambassi}}, \ and\ \bibinfo {author} {\bibfnamefont {A.}~\bibnamefont
  {Silva}},\ }\href {\doibase 10.1103/PhysRevE.87.052129} {\bibfield  {journal}
  {\bibinfo  {journal} {Phys. Rev. E}\ }\textbf {\bibinfo {volume} {87}},\
  \bibinfo {pages} {052129} (\bibinfo {year} {2013})}\BibitemShut {NoStop}%
\bibitem [{\citenamefont {Giorgi}\ \emph {et~al.}(2010)\citenamefont {Giorgi},
  \citenamefont {Paganelli},\ and\ \citenamefont {Galve}}]{giorgi2010}%
  \BibitemOpen
  \bibfield  {author} {\bibinfo {author} {\bibfnamefont {G.}~\bibnamefont
  {Giorgi}}, \bibinfo {author} {\bibfnamefont {S.}~\bibnamefont {Paganelli}}, \
  and\ \bibinfo {author} {\bibfnamefont {F.}~\bibnamefont {Galve}},\
  }\href@noop {} {\bibfield  {journal} {\bibinfo  {journal} {Phys. Rev. A}\
  }\textbf {\bibinfo {volume} {81}},\ \bibinfo {pages} {052118} (\bibinfo
  {year} {2010})}\BibitemShut {NoStop}%
\end{thebibliography}%

\end{document}